\documentclass[a4paper,fleqn,usenatbib]{mnras}

\usepackage[T1]{fontenc}
\usepackage{ae,aecompl}

\usepackage{graphicx}	
\usepackage{amsmath}	
\usepackage{amssymb}	
\usepackage{bm}
\usepackage{xcolor}
\usepackage{xspace}
\usepackage[normalem]{ulem}

\newcommand{\vektor}[1]{\mathbfit{#1}}
\newcommand{\tensor}[1]{\mathbfss{#1}}
\newcommand{\vnabla}{\bm{\nabla}}
\newcommand{\bcdot}{\bm{\cdot}}
\newcommand{\dd}{\mathrm{d}}
\newcommand{\slope}{q}
\newcommand{\cri}{{\mathrm{cr},i}}
\newcommand{\cra}{{\mathrm{cr}}}

\newcommand{\gpcc}{\mathrm{g\,cm^{-3}}}
\newcommand{\kmpersec}{\mathrm{km\,s^{-1}}}
\newcommand{\diffunit}{\mathrm{cm^2\,s^{-1}}}

\newcommand{\GeVc}{\mathrm{GeV}~c^{-1}}
\newcommand{\TeVc}{\mathrm{TeV}~c^{-1}}
\newcommand{\Msol}{\mathrm{M}_\odot}
\newcommand{\Gyr}{\mathrm{Gyr}}
\newcommand{\Myr}{\mathrm{Myr}}

\newcommand{\kpc}{\mathrm{kpc}}

\newcommand{\rkl}[1]{\left(#1\right)}
\newcommand{\ekl}[1]{\left[#1\right]}

\newcommand{\p}{\rmn{p}}
\newcommand{\bvel}{\ensuremath{\bm{v}}}
\newcommand{\uppi}{\ensuremath{\pi}}

\defcitealias{GirichidisEtAl2020}{Paper I}
\defcitealias{GirichidisEtAl2022}{Paper II}

\usepackage[normalem]{ulem} 

\title[Spectral CR-MHD - III]{Spectrally resolved cosmic rays - III. Dynamical impact and properties of the circumgalactic medium}

\author[Philipp Girichidis et al.]{%
Philipp~Girichidis$^{1}$\thanks{E-mail: philipp@girichidis.com},
  Maria~Werhahn$^{2,3}$,
  Christoph~Pfrommer$^{2}$, R\"{u}diger Pakmor$^3$, Volker Springel$^3$
\vspace{0.1cm}\\%
$^{1}$Universit\"{a}t Heidelberg, Zentrum f\"{u}r Astronomie, Institut f\"{u}r Theoretische Astrophysik, Albert-Ueberle-Str. 2, D-69120 Heidelberg, Germany\\%
$^{2}$Leibniz-Institut f\"{u}r Astrophysik (AIP), An der Sternwarte 16, 14482 Potsdam, Germany\\%
$^{3}$Max-Planck-Institut f\"{u}r Astrophysik, Karl-Schwarzschild-Str. 1, 85741 Garching, Germany
}

\date{Accepted XXX. Received YYY; in original form ZZZ}

\pubyear{2022}

\begin{document}
\label{firstpage}
\pagerange{\pageref{firstpage}--\pageref{lastpage}}
\maketitle

\begin{abstract}
Cosmic rays (CRs) are dynamically important in the evolution of galaxies by regulating star formation and powering galactic outflows. However, to what extent CRs regulate galaxy formation depends on the coupling strength of CRs with the ambient plasma and the effective CR transport speed. Moreover, both properties sensitively depend on the CR momentum, which is largely unexplored in three-dimensional hydrodynamical simulations. We perform magneto-hydrodynamical simulations of entire galaxies with masses ranging from $10^{10}$ to $10^{12}\,\Msol$ and compare dynamically coupled CRs in the grey approximation with a spectrally resolved model that includes CR momenta from $0.1\,\GeVc$ to $100\,\TeVc$. We find that hadronic cooling of CRs dominates over Alfv\'{e}n cooling, with the latter emulating CR losses as a result of streaming of CRs down their pressure gradient. While star formation rates and galaxy morphologies are only mildly affected by the spectral CR modelling, mass loading factors of galactic outflows can differ by up to a factor of four in dwarf galaxies. All simulated low-mass halos ($M=10^{10}$, $10^{11}$, and $3\times10^{11}\,\Msol$) drive strong outflows, where CR transport is temporally dominated by advection. In contrast, the Milky Way-mass galaxy with $M=10^{12}\,\Msol$ does not drive sustained outflows, so that CR transport is entirely dominated by diffusion. The effective energy weighted diffusion coefficients vary by two orders of magnitude from the canonical energy-weighted values of $\langle{D}\rangle_{e_\mathrm{cr}}\sim10^{28}\,\diffunit$ in the disc up to $3\times10^{29}\,\diffunit$ in the circumgalactic medium, where we observe substantial temperature and CR pressure differences between our grey and spectral CR models.
\end{abstract}




\section{Introduction}

CRs constitute an important energy reservoir in the interstellar medium of star forming galaxies \citep{StrongMoskalenkoPtuskin2007, GrenierBlackStrong2015,Zweibel2017}. Their energy density is comparable to that in the other components (i.e., the thermal, magnetic, and turbulent energy densities) of the interstellar medium \citep[ISM,][]{Cox2005, NaabOstriker2017}, which enables them to power dynamically relevant processes \citep[see][for a review]{RuszkowskiPfrommer2023}. CRs are known to regulate star formation in the densest regions of the ISM via CR heating and ionization \citep[e.g.,][]{IvlevEtAl2018,PhanMorlinoGabici2018,PadovaniEtAl2020} and drive powerful winds from the ISM into the halo \citep[e.g.,][]{NaabOstriker2017, VeilleuxEtAl2020}. They also constitute a significant fraction of the total pressure in the circum-galactic medium \citep[CGM,][]{Salem2016,BuckEtAl2020,Ji2020}.

Many numerical and analytical studies have shown that CRs are important agents in galaxies and the ISM. Early models investigate CR driven instabilities similar to the Parker instability \citep{Parker1966,HanaszLesch2003}. A simplified advection-diffusion approximation for CR transport reveals their ability to drive galactic winds in one-dimensional analytic calculations \citep{Ipavich1975, ZirakashviliEtAl1996, PtuskinEtAl1997} and numerical models \citep{Breitschwerdt1991,Everett2008,DorfiBreitschwerdt2012, RecchiaBlasiMorlino2016}. Furthermore, three-dimensional simulations of galaxies with dynamically coupled CRs demonstrate that CRs drive powerful outflows in purely hydrodynamical setups \citep[e.g.,][]{JubelgasEtAl2008,UhligEtAl2012,SalemBryan2014, BoothEtAl2013, WienerEtAl2017,SemenovKravtsovCaprioli2021} as well as magneto-hydrodynamical (MHD) models \citep{YangEtAl2012,HanaszEtAl2013,PakmorEtAl2016,RuszkowskiYangZweibel2017,PfrommerEtAl2017,JacobEtAl2018,ChanEtAl2019,DashyanDubois2020,HopkinsEtAl2021,PeschkenEtAl2021}.

CRs have also been included in smaller scale simulations of the ISM, which explicitly resolve the multi-phase nature of the gas and resolve the dynamical launching of the outflow from the disc more accurately than the galactic models \citep{GirichidisEtAl2016a, SimpsonEtAl2016, GirichidisEtAl2018a, FarberEtAl2018, CommerconMarcowithDubois2019, ArmillottaOstrikerJiang2021, RathjenEtAl2021,SimpsonEtAl2022}. These small-scale models reinforce the conclusion that CRs can have an important dynamical impact on the ISM and are crucial for launching and accelerating galactic outflows.

Whether CRs are most effective in driving the warm or hot phase of the ISM, and for which galaxies at which cosmic time CRs have their largest impact, sensitively depends on the CR transport mechanisms and their coupling with the ambient plasma, namely gyro-resonant CR-wave interactions. While CRs below $\sim100\,\mathrm{GeV}$ are expected to scatter off resonant MHD waves generated predominantly by the gyro-resonant (streaming) instability \citep{KulsrudPearce1969} or the intermediate-scale instability \citep{Shalaby2021,ShalabyEtAl2022, ShalabyEtAl2023}, CRs of higher energies are expected to scatter off MHD waves driven primarily by extrinsic turbulence \citep{AloisioBlasiSerpico2015,Zweibel2017,EvoliEtAl2018}. In addition to the gyro-resonant (streaming) instability, CRs can amplify magnetic fields via the non-resonant instability \citep{Bell2004}, which can then reduce the effective transport of CRs on scales close to their sites of acceleration (which are unresolved in the present work). Describing CR transport by a two-moment formulation of CR hydrodynamics enables a self-consistent computation of the macroscopic CR transport speed and the associated diffusion coefficient, using either a steady-state approximation for the resonant wave energy \citep[e.g.,][]{JiangOh2018,HopkinsEtAl2021}, or a formulation that dynamically follows the resonant Alfv\'en wave energy in response to the gyroresonant CR streaming instability and various collisionless damping processes \citep{ThomasPfrommer2019,ThomasPfrommer2022,ThomasEtAl2020,ThomasEtAl2021}.

Furthermore, most CR cooling and transport effects are strong functions of the CR energy. Because the distribution of CRs follows a power law energy spectrum ranging from MeV to $10^{20}$~eV, their interactions with the ambient plasma vary strongly depending on the energy regime. Low-energy CRs ($\lesssim0.1$~GeV) have an increased cross section with the thermal gas and provide most of the heating in dense star-forming cores \citep[e.g.,][]{GirichidisEtAl2020, PadovaniEtAl2020}. CRs in the GeV energy range cool less efficiently. As a result, they are primarily responsible for carrying the pressure of the entire population for realistic values of the momentum spectral index of 2.2 to 2.8. Consequently, they are responsible for driving galactic outflows. While the high-energy part of the CR spectrum with energies $\gtrsim\rmn{TeV}$ is typically subdominant in total energy and therefore dynamically not relevant, these high-energy CRs produce $\gamma$ rays, which is of prime interest for imaging air and water Cerenkov telescopes such as CTA, HAWK, and LHASSO.

In particular, CR ions interact hadronically with the ambient medium and produce pions that decay into $\gamma$ rays, neutrinos, and secondary electrons/positrons. Similarly to primary, shock accelerated electrons, these secondary leptons produce radio synchrotron emission and inverse Compton $\gamma$-ray emission, which enables studying the underlying CR populations via their non-thermal multi-frequency spectra from the radio to PeV $\gamma$-ray range \citep{KoteraOlinto2011,WerhahnEtAl2021a}. A testimony of this is the far-infrared--$\gamma$-ray relation at GeV energies \citep{2012AckermannGamma, 2016RojasBravo,2020Ajello} that has been modelled in three-dimensional MHD simulations \citep{PfrommerEtAl2017b,ChanEtAl2019,WerhahnEtAl2021b,BuckEtAl2020,NunezCastineyraEtAl2023}, indicating that it allows probing the CR ion population via the dominant pion-decay $\gamma$-ray emission, and indirectly its ability to provide dynamical feedback through the calorimetric fraction (i.e., the energy fraction that is radiated away via hadronic CR interactions and thus not available any more for feedback). Similarly interesting is the far-infrared--radio relation \citep{1971VanDerKruit,2003Bell}, which enables probing CR electron calorimetry as well as transport and ageing processes \citep{Thompson2006,Lacki2010,WerhahnEtAl2021c,PfrommerEtAl2022}.

Most previous hydrodynamical models only include CRs in the grey approximation, i.e., they follow the momentum-integrated total CR energy. The transport coefficients as well as the cooling processes are treated in an effective manner without any momentum dependence while assuming a globally universal steady-state CR spectrum \citep[e.g.,][]{PfrommerEtAl2017}. However, numerical simulations that include a full CR spectrum (\citealt{GirichidisEtAl2020}, hereafter Paper I) suggest that the shape of the spectra differ perceptibly in different regions of the galaxy as well as over time (\citealt{GirichidisEtAl2022}, hereafter Paper II). This finding is also supported by a spectral analysis in post-processing of CR-MHD models \citep{WerhahnEtAl2021a}.

In this study, we extend the previous models of spectrally resolved, dynamically coupled CRs \citepalias{GirichidisEtAl2020,GirichidisEtAl2022} in galaxy simulations using the moving mesh code \textsc{Arepo} \citep{Springel2010,WeinbergerSpringelPakmor2020}. We compare spectrally resolved CRs to their grey counterparts and investigate the dynamical and thermal impact of CRs in different halos with masses ranging from dwarfs ($M=10^{10}\,\Msol$) to Milky Way analogues ($M=10^{12}\,\Msol$). We emphasize that the spectral model in \citepalias{GirichidisEtAl2022} erroneously did not account for CR cooling. The old model therefore marks an extreme case of energy dependent CR transport, which we extend here to also include CR cooling.

The outline of the paper is as follows. In Section~\ref{sec:numerics} we briefly summarize the basics of the spectral model \citepalias{GirichidisEtAl2020,GirichidisEtAl2022}. Section~\ref{sec:galaxy-setups} describes the galaxy setups and details the simulations performed. An analysis of the cooling processes is presented in Section~\ref{sec:cooling}, followed by an analysis of the star formation rate and the outflows in Section~\ref{sec:sfr-outflows}. We discuss the effective diffusion coefficients and the dominant transport mechanisms in Sections~\ref{sec:diff-coeff} and \ref{sec:adv-vs-diff}, respectively. The time averaged accelerations and their connection to outflows are discussed in Section~\ref{sec:vertical-force-profiles}. Finally, we explore CGM properties in Section~\ref{sec:cgm-properties}, before finishing the paper with a discussion (Section~\ref{sec:discussion}) and conclusions (Section~\ref{sec:conclusions}). In Appendix~\ref{sec:app-gamma-effective}, we show the effective adiabatic index and the associated pressure of the CR distribution and estimate the error on the gas dynamics associated with our simplified Alfv\'en cooling approach in Appendix~\ref{sec:spectra}. In Appendix~\ref{sec:B_morphology}, we show the morphology of our dynamo-grown magnetic field.

\begin{table*}
  \begin{minipage}{0.7\textwidth}
\caption{List of simulations. Shown are the simulation name, the halo mass $M$, and whether the simulation includes spectral CRs (in this case, $N_\mathrm{bins}$ denotes the number of spectral bins). The next two columns list the diffusion coefficient along the magnetic field, which corresponds to $D(1\,\GeVc)$ in the spectral case, and the momentum scaling index that is defined via $D\propto p^\delta$. The last column indicates whether Alfv\'{e}n cooling is included or not.} 
\begin{tabular}{llcccccc}
\hline
no. & name & mass & spectral & $N_\mathrm{bins}$ & $D_0$           & $\delta$ & Alfv\'{e}n \\
    &      & ($\Msol$) &          &                   &  ($10^{28}$~cm$^2$ s$^{-1}$) &     &        cooling \\
\hline
1  & $M10^{10}\text{-}\mathrm{noCR}$ & $1\times10^{10}$ & -- & -- & -- & -- & no\\
2  & $M10^{10}\text{-}\mathrm{grey}10^{28}\text{-}\Lambda_\mathrm{A}\!=\!0$ & $1\times10^{10}$ & no  & -- & 1  & --  & no \\
3  & $M10^{10}\text{-}\mathrm{grey}10^{28}\text{-}\Lambda_\mathrm{A}$ & $1\times10^{10}$ & no  & -- & 1  & --  & yes\\
4  & $M10^{10}\text{-}\mathrm{grey}4\!\times\!10^{28}\text{-}\Lambda_\mathrm{A}$ & $1\times10^{10}$ & no  & -- & 4  & --  & yes\\
5  & $M10^{10}\text{-}\mathrm{grey}10^{29}\text{-}\Lambda_\mathrm{A}$ & $1\times10^{10}$ & no  & -- & 10  & --  & yes\\
6  & $M10^{10}\text{-}\mathrm{grey}3\!\times\!10^{29}\text{-}\Lambda_\mathrm{A}$ & $1\times10^{10}$ & no  & -- & 30  & --  & yes\\
7  & $M10^{10}\text{-}\mathrm{spec}\text{-}\Lambda_\mathrm{A}$ & $1\times10^{10}$ & yes & 12 & 1  & 0.3 & yes\\
\hline
8  & $M10^{11}\text{-}\mathrm{noCR}$ & $1\times10^{11}$ & -- & -- & -- & -- & no\\
9  & $M10^{11}\text{-}\mathrm{grey}10^{28}\text{-}\Lambda_\mathrm{A}\!=\!0$ & $1\times10^{11}$ & no  & -- & 1 & --  & no\\
10  & $M10^{11}\text{-}\mathrm{grey}10^{28}\text{-}\Lambda_\mathrm{A}$ & $1\times10^{11}$ & no &  -- & 1  & -- & yes\\
11  & $M10^{11}\text{-}\mathrm{grey}4\!\times\!10^{28}\text{-}\Lambda_\mathrm{A}$ & $1\times10^{11}$ & no  & -- & 4  & --  & yes\\
12  & $M10^{11}\text{-}\mathrm{grey}10^{29}\text{-}\Lambda_\mathrm{A}$ & $1\times10^{11}$ & no &  -- & 10  & -- & yes\\
13  & $M10^{11}\text{-}\mathrm{grey}3\!\times\!10^{29}\text{-}\Lambda_\mathrm{A}$ & $1\times10^{11}$ & no  & -- & 30  & --  & yes\\
14  & $M10^{11}\text{-}\mathrm{spec}\text{-}\Lambda_\mathrm{A}\!=\!0$ & $1\times10^{11}$ & yes & 12 & 1  & 0.3 & no \\
15  & $M10^{11}\text{-}\mathrm{spec}\text{-}\Lambda_\mathrm{A}$ & $1\times10^{11}$ & yes & 12 & 1  & 0.3 & yes\\
16  & $M10^{11}\text{-}\mathrm{spec}\text{-}\delta0.5\text{-}\Lambda_\mathrm{A}$ & $1\times10^{11}$ & yes & 12 & 1  & 0.5 & yes \\
\hline
17  & $M3\!\times\!10^{11}\text{-}\mathrm{noCR}$ & $3\times10^{11}$ & -- & -- & -- & -- & no\\
18  & $M3\!\times\!10^{11}\text{-}\mathrm{grey}10^{28}\text{-}\Lambda_\mathrm{A}\!=\!0$ & $3\times10^{11}$ & no  & -- & 1  & --  & no \\
19  & $M3\!\times\!10^{11}\text{-}\mathrm{grey}10^{28}\text{-}\Lambda_\mathrm{A}$ & $3\times10^{11}$ & no  & -- & 1  & --  & yes \\
20  & $M3\!\times\!10^{11}\text{-}\mathrm{grey}4\!\times\!10^{28}\text{-}\Lambda_\mathrm{A}$ & $3\times10^{11}$ & no  & -- & 4  & --  & yes\\
21  & $M3\!\times\!10^{11}\text{-}\mathrm{grey}10^{29}\text{-}\Lambda_\mathrm{A}$ & $3\times10^{11}$ & no  & -- & 10  & --  & yes \\
22  & $M3\!\times\!10^{11}\text{-}\mathrm{grey}3\!\times\!10^{29}\text{-}\Lambda_\mathrm{A}$ & $3\times10^{11}$ & no  & -- & 30  & --  & yes\\
23  & $M3\!\times\!10^{11}\text{-}\mathrm{spec}\text{-}\Lambda_\mathrm{A}$ & $3\times10^{11}$ & yes & 12 & 1  & 0.3 & yes\\
\hline
24 & $M10^{12}\text{-}\mathrm{noCR}$ & $1\times10^{12}$ & -- & -- & -- & -- & no\\
25  & $M10^{12}\text{-}\mathrm{grey}10^{28}\text{-}\Lambda_\mathrm{A}\!=\!0$ & $1\times10^{12}$ & no  & -- & 1  & --  & no \\
26  & $M10^{12}\text{-}\mathrm{grey}10^{28}\text{-}\Lambda_\mathrm{A}$ & $1\times10^{12}$ & no  & -- & 1  & --  & yes\\
27  & $M10^{12}\text{-}\mathrm{grey}4\!\times\!10^{28}\text{-}\Lambda_\mathrm{A}$ & $1\times10^{12}$ & no  & -- & 4  & --  & yes\\
28  & $M10^{12}\text{-}\mathrm{grey}10^{29}\text{-}\Lambda_\mathrm{A}$ & $1\times10^{12}$ & no  & -- & 10  & --  & yes\\
29  & $M10^{12}\text{-}\mathrm{grey}3\!\times\!10^{29}\text{-}\Lambda_\mathrm{A}$ & $1\times10^{12}$ & no  & -- & 30  & --  & yes\\
30  & $M10^{12}\text{-}\mathrm{spec}\text{-}\Lambda_\mathrm{A}$ & $1\times10^{12}$ & yes & 12 & 1  & 0.3 & yes\\
31  & $M10^{12}\text{-}\mathrm{spec}\text{-}\delta0.4\text{-}\Lambda_\mathrm{A}$ & $1\times10^{12}$ & yes & 12 & 1  & 0.4 & yes\\
32  & $M10^{12}\text{-}\mathrm{spec}\text{-}\delta0.5\text{-}\Lambda_\mathrm{A}$ & $1\times10^{12}$ & yes & 12 & 1  & 0.5 & yes\\
\hline
\end{tabular}
  \end{minipage}
  \label{tab:simulations}
\end{table*}

\begin{table}
\caption{Spectral bins and the corresponding diffusion coefficients. Listed are the bin number, the left-hand momentum boundary, the central momentum as well as the corresponding diffusion coefficient for three different scalings of $D$ with the momentum.}
\label{tab:spec-bins}
\begin{tabular}{cccccc}
\hline
& & & \multicolumn{3}{c}{$D(p)$}\\
$i$ & $p_{i-1/2}$ & $p_i$ & $\delta=0.3$ & $\delta=0.4$ & $\delta=0.5$ \\
  & (GeV/$c$) & (GeV/$c$) & \multicolumn{3}{c}{($10^{28}$\,cm$^2$/s)}\\
  \hline
 0   &   $0.1$             &   0.178          & 0.60 & 0.50 & 0.42       \\
 1   &   $0.316$           &   0.562          & 0.84 & 0.79 & 0.75     \\
 2   &   $1.0$             &   1.78           & 1.19 & 1.26 & 1.33     \\
 3   &   $3.16$            &   5.62           & 1.68 & 2.0 & 2.37      \\
 4   &   $10.0$            &  17.8            & 2.37 & 3.16 & 4.22     \\
 5   &   $31.6$            &  56.2            & 3.35 & 5.01 & 7.5      \\
 6   &   $100$             & 178              & 4.73 & 7.94 & 13.34    \\
 7   &   $316$             & 562              & 6.68 & 12.59 & 23.71   \\
 8   &   $10^3$            & $1.78\times10^3$ & 9.44 & 19.95 & 42.17   \\
 9   &   $3.16\times10^3$  & $5.62\times10^3$ & 13.34 & 31.62 & 74.99  \\
10   &   $10^4$            & $1.78\times10^4$ & 18.84 & 50.12 & 133.35 \\
11   &   $3.16\times10^4$  & $5.62\times10^4$ & 26.61 & 79.43 & 237.14 \\
\hline
\end{tabular}
\end{table}

\section{Numerical Methods}
\label{sec:numerics}

\subsection{Spectral CR description and coupling to MHD}
We use the spectrally resolved CR-MHD model developed in \citetalias{GirichidisEtAl2020} (spectral scheme) and \citetalias{GirichidisEtAl2022} (coupling to MHD), of which we highlight the main concept and features here. The numerical model is based on solving the Fokker-Planck equation for CRs (see \citealt{Skilling1975} for a theoretical derivation, and \citealt{Miniati2001, YangRuszkowski2017, OgrodnikHanaszWoltanski2021} for recent numerical solutions as well as 
\citealt{HanaszStrongGirichidis2021} for a recent review on numerical CR models),
\begin{align}
  \frac{\partial f}{\partial t} = & \underbrace{-\bvel\bcdot\vnabla f}_{\text{advection}} + \underbrace{\vnabla\bcdot\rkl{\tensor{D}_{xx}\bcdot \vnabla f}}_{\text{diffusion}} + \underbrace{\frac 1 3 \rkl{\vnabla\bcdot\bvel}p\frac{\partial f}{\partial p}}_{\text{adiabatic process}}\nonumber\\
  & + \underbrace{\frac{1}{p^2}\frac{\partial}{\partial p}\ekl{p^2\rkl{b_l f + D_{pp}\frac{\partial f}{\partial p}}}}_{\text{other losses and Fermi II acceleration}} + \underbrace{j}_{\text{sources}}\label{eq:FP},
\end{align}
where $f=f(\vektor{x},p,t)=\dd^6 N/(\dd x^3\,\dd p^3)$ denotes the isotropic part of the CR distribution function, with the spatial coordinates $\vektor{x}$, the momentum $p=|\vektor{p}|$, and the time $t$. The mean velocity of the gas is $\bvel$, the spatial diffusion tensor is $\tensor{D}_{xx}=\tensor{D}_{xx}(\vektor{x},p,t)$, and the diffusion coefficient in momentum space is $D_{pp}=D_{pp}(\vektor{x},p,t)$. CR losses and sources are expressed as $b_l=b_l(\vektor{x},\vektor{p},t)=\dd p/\dd t$ and $j=j(\vektor{x},\vektor{p},t)$, respectively. The losses $b_l$ include Coulomb and hadronic interactions. Since we neglect diffusion in momentum space ($D_{pp}=0$), we will omit the subscript of the spatial diffusion coefficient in the following.

The momentum space is discretised using $N_\mathrm{spec}$ logarithmic bins with bin face momenta $p_{i-1/2}$ and bin centred counterparts $p_i = \sqrt{p_{i-1/2}p_{i+1/2}}$. The particle distribution function $f$ is described by piecewise power laws 
\begin{align}
  f(p) &= \sum_{i=1}^{N_\mathrm{spec}} f_i(p) \\
  &= \sum_{i=1}^{N_\mathrm{spec}} f_{i-1/2} \rkl{\frac{p}{p_{i-1/2}}}^{-\slope_i}
  \theta\rkl{p-p_{i-1/2}}\theta\rkl{p_{i+1/2}-p}
\end{align}
where $p_{i-1/2}$ is the momentum and $f_{i-1/2}$ is the corresponding amplitude at the left-hand face of momentum bin $i$. The slope in bin $i$ is denoted by $\slope_i$ and $\theta$ is the Heaviside function. We note that the spectrum does not need to be continuous at the bin faces, which increases numerical accuracy and avoids instabilities \citepalias[see][]{GirichidisEtAl2020}. Solving for the two degrees of freedom per bin ($f_{i-1/2}$ and $\slope_i$) requires solving for two independent quantities related to $f$. We choose the number and energy density,
\begin{align}
\label{eq:number-density}
  n_i &= \int_{p_{i-1/2}}^{p_{i+1/2}} 4\pi p^2 f(p) \,\dd p,\\
\label{eq:energy-density}
  e_i &= \int_{p_{i-1/2}}^{p_{i+1/2}} 4\pi p^2 f(p) T(p)\,\dd p,
\end{align}
where $T(p)=\sqrt{p^2c^2 + m_\mathrm{p}^2c^4} - m_\mathrm{p} c^2$ is the kinetic energy of individual protons and $c$ is the speed of light. The CRs are coupled to the MHD equations via the gas velocity $\bvel$ for advection, and the total CR pressure
\begin{align}
P_\cra &= \sum_{i=1}^{N_\mathrm{spec}} P_\cri= \frac{4\pi}{3}\,\sum_{i=1}^{N_\mathrm{spec}}\int_{p_{i-1/2}}^{p_{i+1/2}}\,\frac{f_i(p)\,p^4c^2}{\sqrt{m_\p^2c^4+p^2c^2}}\dd p
\end{align}
for the dynamical coupling in the Riemann solver \citep[see][]{PfrommerEtAl2017}.

\subsection{Spatial diffusion, streaming approximation and cooling}

We use momentum dependent anisotropic diffusion along the direction of the magnetic field using the implicit solver described in \citet{Pakmor2016}. An improvement in terms of spectral stability \citepalias{GirichidisEtAl2022} allows us to only diffuse the CR energy density -- rather than both number and energy density -- and reconstruct the spectrum in a stable and energy conserving manner after the diffusion step. Diffusion is applied to each spectral bin with a bin-centred diffusion coefficient parallel to the magnetic field\footnote{All quoted values in the paper correspond to parallel diffusion coefficients and in the following, we drop the subscript $\parallel$ for simplicity.}
\begin{equation}
\label{eq:diff-p-scaling}
D(p_i) = D_{\parallel,0}\,\rkl{\frac{p_i}{1\,\mathrm{GeV}~c^{-1}}}^\delta.
\end{equation}
Diffusion itself is an energy conserving process, and in order to emulate CR streaming along the magnetic field, we additionally account for CR Alfv\'en wave losses. The streaming instability effectively reduces the mean CR transport speed, $\bvel_{\mathrm{st}}$, to the Alfv\'{e}n speed,
\begin{equation}
\bvel_{\mathrm{st}}=-\bvel_{\mathrm{A}} \operatorname{sgn}\left(\boldsymbol{B} \bm{\cdot} \bm{\nabla} {P}_{\mathrm{cr}}\right)=-\frac{\boldsymbol{B}}{\sqrt{4 \pi \rho}} \frac{\boldsymbol{B} \bm{\cdot} \bm{\nabla} {P}_{\mathrm{cr}}}{\left|\boldsymbol{B} \bm{\cdot} \bm{\nabla} {P}_{\mathrm{cr}}\right|},
\end{equation}
where $\boldsymbol{B}$ denotes the magnetic field and $\rho$ is the gas mass density (assuming full ionisation). Instead of a self-consistent treatment of the CR transport speed and losses, as obtained in recent two-moment descriptions of CR hydrodynamics \citep[see e.g.,][]{JiangOh2018,ThomasPfrommer2019,HopkinsEtAl2020,DashyanDubois2020,HopkinsEtAl2021,ArmillottaOstrikerJiang2021,ThomasPfrommer2022}, we apply an energy conserving diffusion step in combination with effective CR losses due to resonant excitation of Alfv\'{e}n waves as derived in \citet{WienerZweibelOh2013},
\begin{equation}
    \label{eq:alfv-cool}
    \Lambda_\mathrm{A} = \bvel_\mathrm{A}\bcdot\vnabla P_\mathrm{cr}.
\end{equation}
We apply $\Lambda_{A}$ to the total CR energy after the diffusion step and rescale the spectrum to match the new CR energy. In Appendix~\ref{sec:spectra}, we show our simulated CR spectra and demonstrate that the error on the gas dynamics associated with this rescaling is expected to be small. In addition, we account for Coulomb, ionization and hadronic losses as described in \citetalias{GirichidisEtAl2020}.

\section{Galaxy simulation setup}
\label{sec:galaxy-setups}

We use the \textsc{Arepo} code, which employs an unstructured moving mesh to decompose the computational domain. The hydrodynamical solver is second-order accurate in space and time \citep{Springel2010,Pakmor2016a,WeinbergerSpringelPakmor2020}. CRs are dynamically coupled to the evolution of the gas as described in \citet{PfrommerEtAl2017} using the advection-diffusion approximation. The spectral solver for the CRs is coupled to the MHD and the diffusion solver as outlined in \citetalias{GirichidisEtAl2022}. Magnetic fields are included in the ideal MHD approximation \citep{PakmorBauerSpringel2011,PakmorSpringel2013} using the Powell 8-wave scheme for divergence
cleaning of the magnetic field \citep{PowellEtAl1999}. This enables us to resolve magnetic field growth via a small-scale dynamo already for turbulent injection length scales of order 1~kpc \citep{PakmorEtAl2017,PfrommerEtAl2022}. CR diffusion uses the semi-implicit diffusion solver by \citet{PakmorEtAl2016}.

For each galaxy we simulate, the numerical setup consists of an isolated collapsing gas cloud that is embedded in a dark matter halo, similar to the setups in \citet{PakmorEtAl2016} and \citet{PfrommerEtAl2017}. We adopt a baryon mass fraction of $\Omega_\mathrm{b}/\Omega_\mathrm{m}=0.155$ \citep{Planck2020_VI} for all our models and vary the total mass of the system. Our four different haloes have total masses of $10^{10}\,\mathrm{M}_\odot$, $10^{11}\,\mathrm{M}_\odot$, $3\times10^{11}\,\mathrm{M}_\odot$ and $10^{12}\,\mathrm{M}_\odot$ (our Milky Way-like halo) with corresponding initial gas masses of $1.55\times10^9\,\mathrm{M}_\odot$, $1.55\times10^{10}\,\mathrm{M}_\odot$, $4.65\times10^{10}\,\mathrm{M}_\odot$ and $1.55\times10^{11}\,\mathrm{M}_\odot$, respectively. The dark matter and the gas initially follow an NFW profile \citep{NavarroFrenkWhite1997} with a concentration parameter of $c_{200} = r_{200}/r_\mathrm{s} = 7$. Here, $r_\mathrm{s}$ denotes the characteristic scale radius of the NFW profile. The radius $r_{200}$ encloses a mass with an average density 200 times the critical density of the universe. While we adopt an external fixed dark matter potential, the gas follows initially a solid body rotation with a dimensionless spin parameter $\lambda=J|E|^{1/2}/(GM_{200}^{5/2})=0.3$. Here, $J$ is the angular momentum, $|E|$ the total halo energy and $G$ Newton's constant. The mass inside $r_{200}$ is given by $M_{200}$, which we abbreviate with $M$ in our setups to keep the simulation names short. We set an initial seed field along the $x$-axis with a field strength of $10^{-4}\,\mu\mathrm{G}$. The halo is sampled with $10^6$ gas cells with a uniform target mass resolution of $10^{-6}$ times the total gas mass, which ranges from $1.55\times10^3\,\mathrm{M}_\odot$ for the galaxy with the lowest mass up to $1.55\times10^5\,\mathrm{M}_\odot$ for our Milky Way-like system, respectively. Adjacent cells are only allowed to differ by a factor of ten in volume.

We include radiative cooling and star formation employing the pressurized ISM model by \citet{SpringelHernquist2003}, which includes a stochastic star formation prescription that reproduces the observational relation between the gas surface density and the star formation rate surface density \citep{KennicuttSchmidt1998}. The subgrid ISM model assumes the hot and cold phase of the ISM to be in pressure equilibrium. The total ISM pressure is modeled with a stiff equation of state above a threshold gas density for star formation, $n_\mathrm{SF}= 0.13\; \mathrm{cm}^{-3}$. Star formation is implemented as an injection of a star particle with instantaneous, implicit stellar feedback as the effective equation of state used as subgrid ISM model does not require modelling explicit thermal/kinetic feedback. We inject CRs with an efficiency of 10 per cent of the canonical supernova (SN) injection energy of $10^{51}\,\mathrm{erg}$ \citep[see, e.g.,][]{HelderEtAl2012,MorlinoCaprioli2012,AckermannEtAl2013} with an injection spectrum of $f(p)\propto p^{4.2}$. However, detailed comparisons of numerical models with observations of individual SNe suggest that also lower efficiencies of only 5 per cent are plausible \citep{PaisEtAl2018}. Initially, there are no CRs in the models, so all CR energy is injected by SNe.

We present results from 32 simulations, which are listed in Table~\ref{tab:simulations}. We distinguish between grey CR models similar to the galaxies in \citet{PakmorEtAl2016, PfrommerEtAl2017} using different diffusion coefficients (``$\mathrm{grey}$''), spectral CR models (``$\mathrm{spec}$'') and control runs without CRs (``noCR''). We further differentiate between models with and without Alfv\'{e}n cooling ($\Lambda_\mathrm{A}$ and $\Lambda_\mathrm{A}=0$). For the Milky Way-like galaxies, we furthermore study two different momentum scalings of the diffusion coefficient (Eq.~\eqref{eq:diff-p-scaling}) with $\delta=0.3$ to $0.5$ \citep[see][for a discussion of the momentum scaling]{Evoli2020, WerhahnEtAl2021a, WerhahnEtAl2021b, WerhahnEtAl2021c}. The momenta for the spectral bin edges, the bin centres, and the corresponding bin-centred diffusion coefficients are listed in Table~\ref{tab:spec-bins}. 

For the spectral models we use 12 momentum bins ranging from $0.1\,\mathrm{GeV\,}c^{-1}$ to $100\,\mathrm{TeV\,}c^{-1}$. We inject CRs with a power-law source spectrum, $f(p)\propto p^{-4.2}$, into the local environment of newly created star particles and use a spherical top hat filter containing the closest 32 mesh cells surrounding the star. We run all simulations for a total time of $3\,\mathrm{Gyr}$.

\begin{figure*}
\begin{minipage}{\textwidth}
    \centering
    \includegraphics[width=\textwidth]{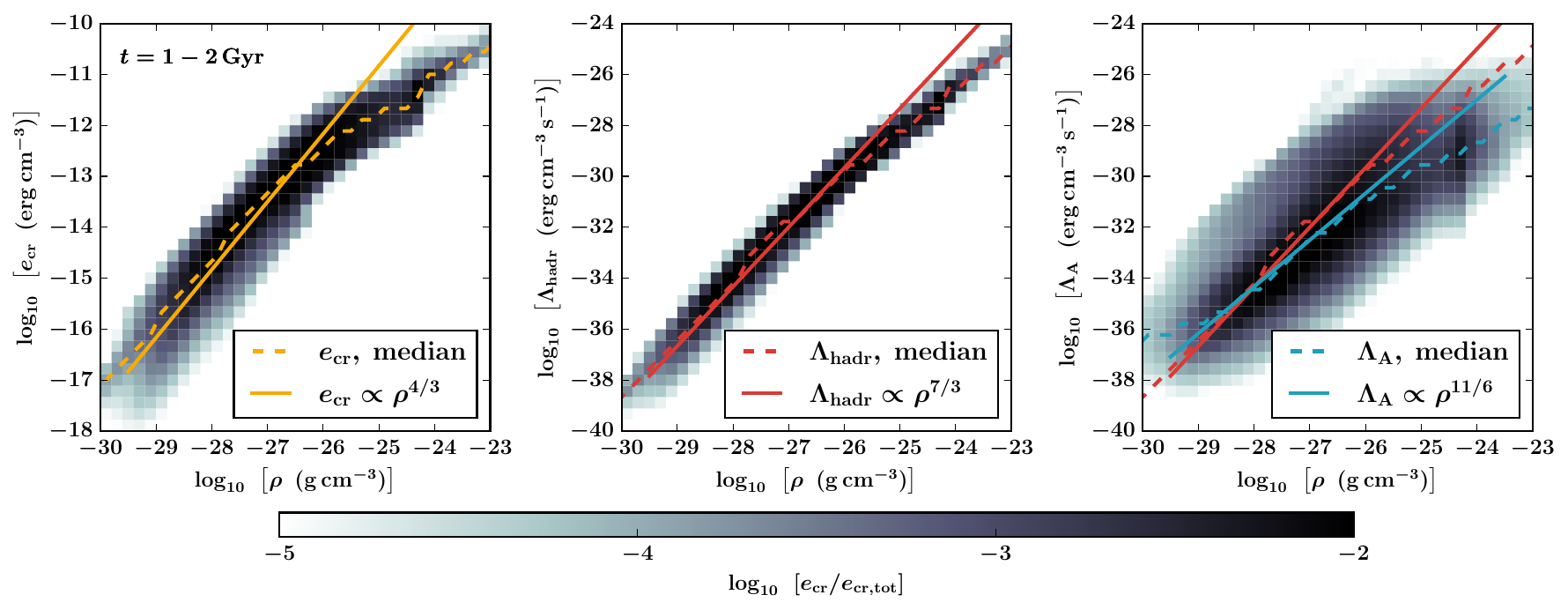}
    \caption{Distribution of the CR energy density, the hadronic and the Alfv\'{e}n cooling rate as a function of density in the simulation ``$M10^{11}\text{-}\mathrm{spec}\text{-}\Lambda_\mathrm{A}$'' averaged over $t=1-2\,\Gyr$. Colour coded is the distribution in these variables, weighted with the fraction of CR energy density. Overlaid are the median (dashed lines) as well as the analytical scaling that assumes adiabatic CRs only (solid lines), which is only valid at low densities. Hadronic cooling dominates over Alfv\'{e}n cooling above a density of $\rho\sim10^{-28}\,\gpcc$. This explains the small differences in the galaxy properties between the simulations with and without Alfv\'{e}n cooling.}
    \label{fig:alfven-cooling-1e11-spec12}
\end{minipage}
\end{figure*}

\begin{figure*}
\begin{minipage}{\textwidth}
    \centering
    \includegraphics[width=\textwidth]{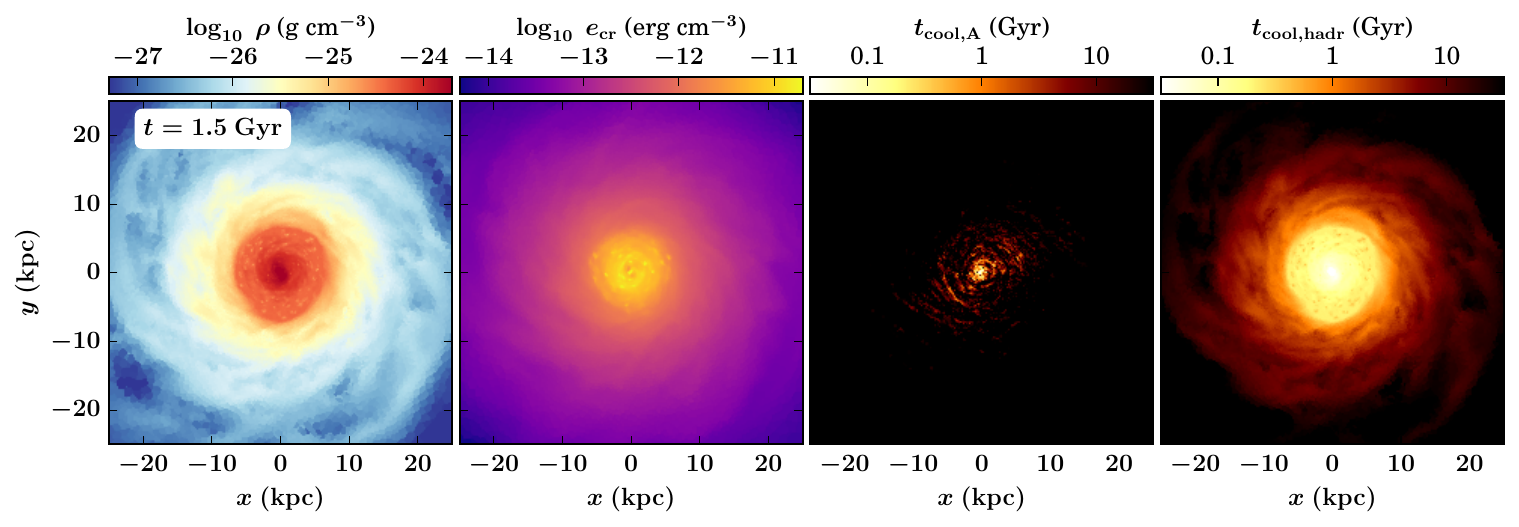}
    \caption{Illustration of the cooling times for simulation ``$M10^{11}\text{-}\mathrm{spec}\text{-}\Lambda_\mathrm{A}$'' at $t=1.5\,\Gyr$. From left to right we show gas density, the total CR energy density, and the cooling times for Alfv\'{e}n cooling and hadronic cooling, respectively. The Alfv\'{e}n cooling time is significantly larger than the hadronic counterpart in most of the volume. Only in the centre, $t_\mathrm{cool,A}$ is comparable to the dynamical time. This is contrary to the hadronic case, where most of the disc has cooling times perceptibly shorter than the dynamical time, which is a necessary condition for CR calorimetry.}
    \label{fig:alfven-cooling-maps}
\end{minipage}
\end{figure*}

\section{CR cooling processes}
\label{sec:cooling}

Before analyzing the simulations we can estimate the expected scalings of the CR energy distribution and the cooling times with gas density. For adiabatic changes of the CRs, we expect the CR energy density and pressure to scale as $e_\mathrm{cr},P_\mathrm{cr}\propto\rho^{4/3}$, where we adopt the adiabatic index $\gamma=4/3$ for a relativistic fluid.\footnote{We compare the spectrally computed adiabatic index to the simple relativistic one in Section~\ref{sec:app-gamma-effective}.} At high gas densities, non-adiabatic CR interactions with the gas as well as active CR transport processes such as diffusion are expected to cause a deviation of this adiabatic scaling \citep[see, e.g., figure~16 of][]{GirichidisEtAl2018a}. For hadronic interactions with ambient nuclei of the ISM, the loss rate of kinetic CR energy reads
\begin{equation}
    -\left(\frac{\dd T}{\dd t}\right)_\mathrm{hadr} = c\,n\,\sigma_\mathrm{pp}K_\mathrm{p}T(p)\theta(p-p_\mathrm{thr}),
\end{equation}
where $c$ denotes the speed of light, $n$ the target nucleon density of the ISM, $\sigma_\mathrm{pp}$ is the total pion cross section and $K_\mathrm{p}\approx 1/2$ represents the inelasticity of the reaction \citep{MannheimSchlickeiser1994}. The threshold momentum above which pions can be produced hadronically is $p_\mathrm{thr}\approx0.78\,\GeVc$ ($\theta$ is the Heaviside step function). The cooling rate thus scales linearly with $n$, and the cooling time $T/\dot{T}$ is independent of the CR momentum\footnote{We note that the cooling time defined via the CR momenta ($t_\mathrm{cool,hadr}' = p/\dot{p}$) has a small momentum dependence and is slightly enhanced for mildly relativistic CRs in comparison to the ultra-relativistic limit, see e.g., figure~2 in \citetalias{GirichidisEtAl2020}.} with
\begin{equation}
\label{eq:t_hadr}
    t_\mathrm{cool,hadr} \approx 66\,\left(\frac{n}{1\,\mathrm{cm}^{-3}}\right)^{-1}\,\Myr.
\end{equation}
Thus, the hadronic cooling rate has the following scaling with gas density for adiabatic CRs:
\begin{equation}
    \Lambda_\mathrm{hadr}\approx\frac{e_\mathrm{cr}}{t_\mathrm{cool,hadr}}\propto \rho^{7/3}.
\end{equation}
In order to derive the scaling of the Alfv\'{e}n cooling rate with gas density, we investigate the individual terms that enter equation~\eqref{eq:alfv-cool}. Assuming magnetic flux-freezing for isotropic changes of the volume, we have $|\vektor{B}|\propto\rho^{2/3}$ in the kinematic limit \citep[e.g.,][]{Spruit2013}, so that $|\bvel_\mathrm{A}|\propto \rho^{1/6}$. We approximate the gradient of the CR pressure as $\vnabla P_\mathrm{cr}\sim P_\mathrm{cr}/L_\mathrm{cr}$, where $L_\mathrm{cr}$ is the gradient length of the CR pressure, and assume $L_\mathrm{cr}\propto\rho^{-1/3}$ for the scaling of the CR gradient length. Assuming adiabatic changes of the CR pressure, the Alfv\'{e}n wave losses are expected to scale as
\begin{equation}
\label{eq:Lambda_A}
  \Lambda_\mathrm{A}=\bvel_\mathrm{A}\bcdot\vnabla P_\mathrm{cr}\propto\rho^{11/6},
\end{equation}
and the corresponding cooling time scales for adiabatic CRs as
\begin{equation}
    t_\mathrm{cool,A} \approx \frac{e_\mathrm{cr}}{\Lambda_\mathrm{A}}\propto \rho^{-1/2},
\end{equation}
which shows a weaker density scaling in comparison to the hadronic CR interactions (cf.\ Eq.~\ref{eq:t_hadr}).

Figure~\ref{fig:alfven-cooling-1e11-spec12} illustrates from left to right the distribution of the CR energy density, the hadronic as well as the Alfv\'{e}n cooling rate as a function of density for simulation ``$M10^{11}\text{-}\mathrm{spec}\text{-}\Lambda_\mathrm{A}$'' averaged from $t=1-2\,\Gyr$. Averaging over different time intervals does not yield any differences \citep[except for the very beginning of the simulation when the magnetic dynamo has not yet saturated,][]{PfrommerEtAl2022}. Furthermore, the other simulations behave similarly. Colour coded is the fraction of the total CR energy.\footnote{For every snapshot that we include in the time-averaged distribution, we first normalise the CR energy by the total CR energy and then collect the cell data from all snapshots. This ensures that each snapshot has an equal weight in the averaged distribution irrespective of its total CR energy content. We then compute the 2D histogram and divide it by the number of snapshots.} The dashed lines show the median along the ordinate, the solid lines indicate the analytic scaling of the corresponding variable with $\rho$ while assuming adiabatic CRs. The scaling of the energy density closely follows the adiabatic approximation for low densities. Above $\rho\sim10^{-26}\,\gpcc$ the scaling starts to flatten, which is related to non-adiabatic effects such as non-adiabatic cooling in the dense gas and diffusion of CRs out of the dense gas, where they have been injected. The relatively tight distribution of $e_\mathrm{cr}$ around the median leads to a narrow distribution of the hadronic cooling rate, which follows the adiabatic CR scaling up to that density, at which point the adiabatic approximation for $e_\mathrm{cr}$ breaks down. 

The comparison of the hadronic with the Alfv\'{e}n cooling rate in the right-hand panel illustrates two important features. First, the distribution shows a significantly larger scatter, which is the result of larger local differences in the magnetic field strength and local CR pressure gradients. Second, it shows that for densities above $\rho\sim10^{-28}\,\gpcc$, the hadronic cooling dominates over the Alfv\'{e}n cooling rate. In the central region of the galaxy at densities above $\rho\gtrsim10^{-25}\,\gpcc$ the median of the hadronic cooling rate is approximately two orders of magnitude larger than the Alfv\'{e}nic counterpart, which explains the small differences between the simulations with and without Alfv\'{e}n cooling.

We illustrate the spatial distribution of the Alfv\'{e}n and hadronic cooling times in Fig.~\ref{fig:alfven-cooling-maps}, where we show the gas density, the CR energy density, the Alfv\'{e}n cooling time and the hadronic cooling time in cuts through the centre of the disc for the simulation ``$M10^{11}\text{-}\mathrm{spec}\text{-}\Lambda_\mathrm{A}$'' at $t=1.5\,\Gyr$. Both the density and the CR energy density show a relatively smooth distribution across three orders of magnitude. The measured cooling times $t_\mathrm{cool,A}$ range from $\sim100\,\Myr$ in the centre to values larger than the Hubble time in the outskirts of the galactic disc. Only in the innermost $\sim5\,\kpc$ the cooling times are comparable or shorter than the dynamcial time of the disc ($\lesssim100\,\Myr$). For the hadronic counterpart ($t_\mathrm{cool,hadr}$) we find cooling times ranging from $\sim10\,\Myr$ to $\sim10\,\Gyr$.  For most of the volume, $t_\mathrm{cool,hadr}\ll t_\mathrm{cool,A}$. Furthermore, for most parts of the disc, the hadronic cooling time is shorter or comparable to the dynamical time of the disc, which is a necessary condition for CR calorimetry.

\begin{figure}
    \centering
    \includegraphics[width=8cm]{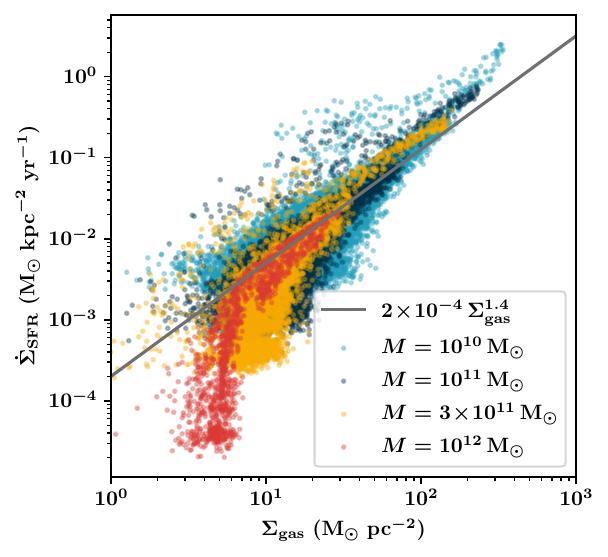}
    \caption{Kennicutt-Schmidt relation for all simulations, colour coded by their halo masses (see legend). The differences between the models at the identical halo mass is negligible. The different halo masses mainly occupy different ranges in the gas column density. Below a surface density of $\Sigma_\mathrm{gas}\approx10\,\mathrm{M}_\odot\,\mathrm{pc}^{-2}$ the star formation rate reduces compared to the classical relation, which is a well-known observed feature \citep[e.g.,][]{BigielEtAl2008,LeroyEtAl2008}}
    \label{fig:KS-relation}
\end{figure}

\begin{figure*}
\begin{minipage}{\textwidth}
    \centering
    \includegraphics[width=\textwidth]{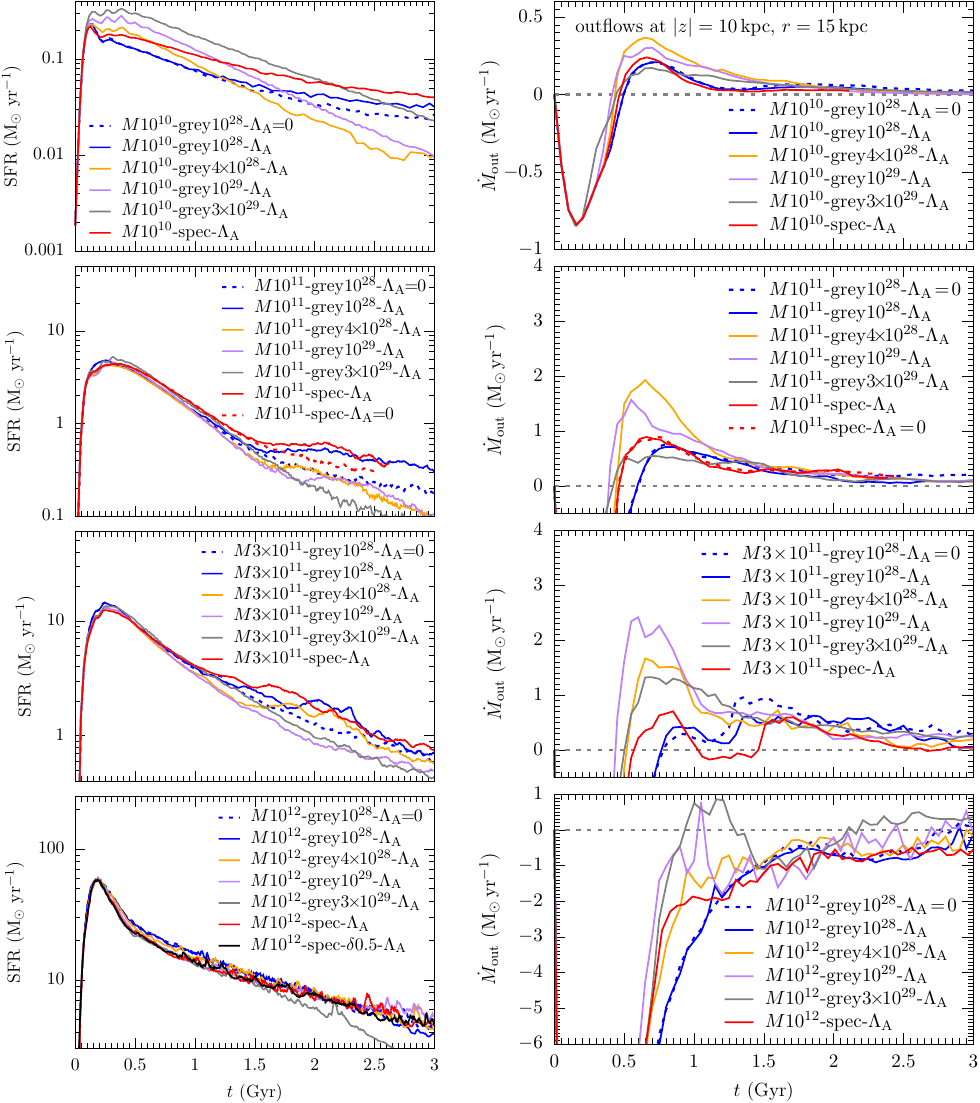}

    \caption{Left: Time evolution of the star formation rate for all halo masses (top to bottom). The peak of the star formation rate hardly differs between the grey and spectral models, except for the halo with the lowest mass ($M=10^{10}\Msol$). At late times, the spectral models show higher star formation rates for the two setups with low halo masses ($M=10^{10}\Msol$ and $M=10^{11}\Msol$) but not for the two more massive ones. Right: Outflow rate as a function of time for all halo masses measured at a height of $10\,\kpc$ above and below the midplane through a circular area with radius $r=15\,\kpc$. We measure noticeable outflows in all but the highest mass halo ($M=10^{12}\,\Msol$), which is dominated by infall. The strongest outflows are driven by the grey models with diffusion coefficients $D=4\times10^{28}$ and $10^{29}\,\diffunit$ with a peak at around $t\sim0.5-1\,\Gyr$. For the most massive halo ($M=10^{12}\,\Msol$) the grey model with the highest diffusion coefficient ($D=3\times10^{29}\,\diffunit$) can drive an outflow for a short period of time.}
    \label{fig:sfr-outflow-time-evol}
\end{minipage}
\end{figure*}

\begin{figure}
    \centering
    \includegraphics[width=8cm]{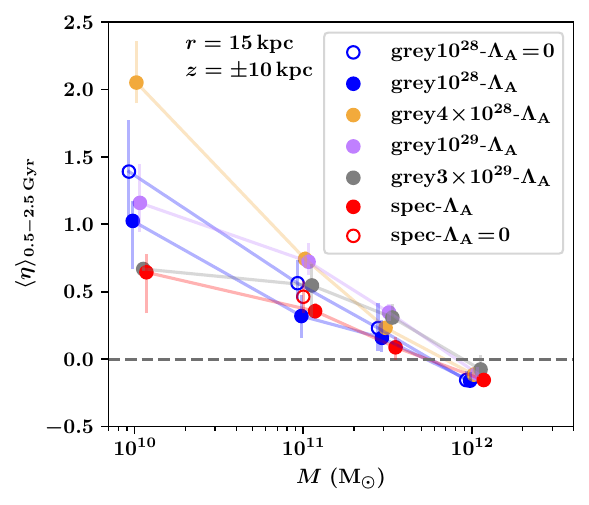}
    \caption{Time-averaged ($0.5-2.5\,\Gyr$) mass loading factor as a function of total mass. The outflow is measured as in Fig.~\ref{fig:sfr-outflow-time-evol} and the mass loading factor is computed for outflow and star formation rates at the same time. All but the highest mass galaxy develop effective outflows with positive mass loading factors. The differences between the CR models is stronger for lower masses. The largest mass loading factors are reached by the grey simulations with $D=4\times10^{28}$ and $10^{29}\,\diffunit$. The temporal scatter is indicated by the errorbars marking the 25 and 75 percentile.}
    \label{fig:mass-loading-factor}
\end{figure}

\section{Star formation and outflow rate}
\label{sec:sfr-outflows}

\begin{figure*}
\begin{minipage}{\textwidth}
    \centering
    \includegraphics[width=\textwidth]{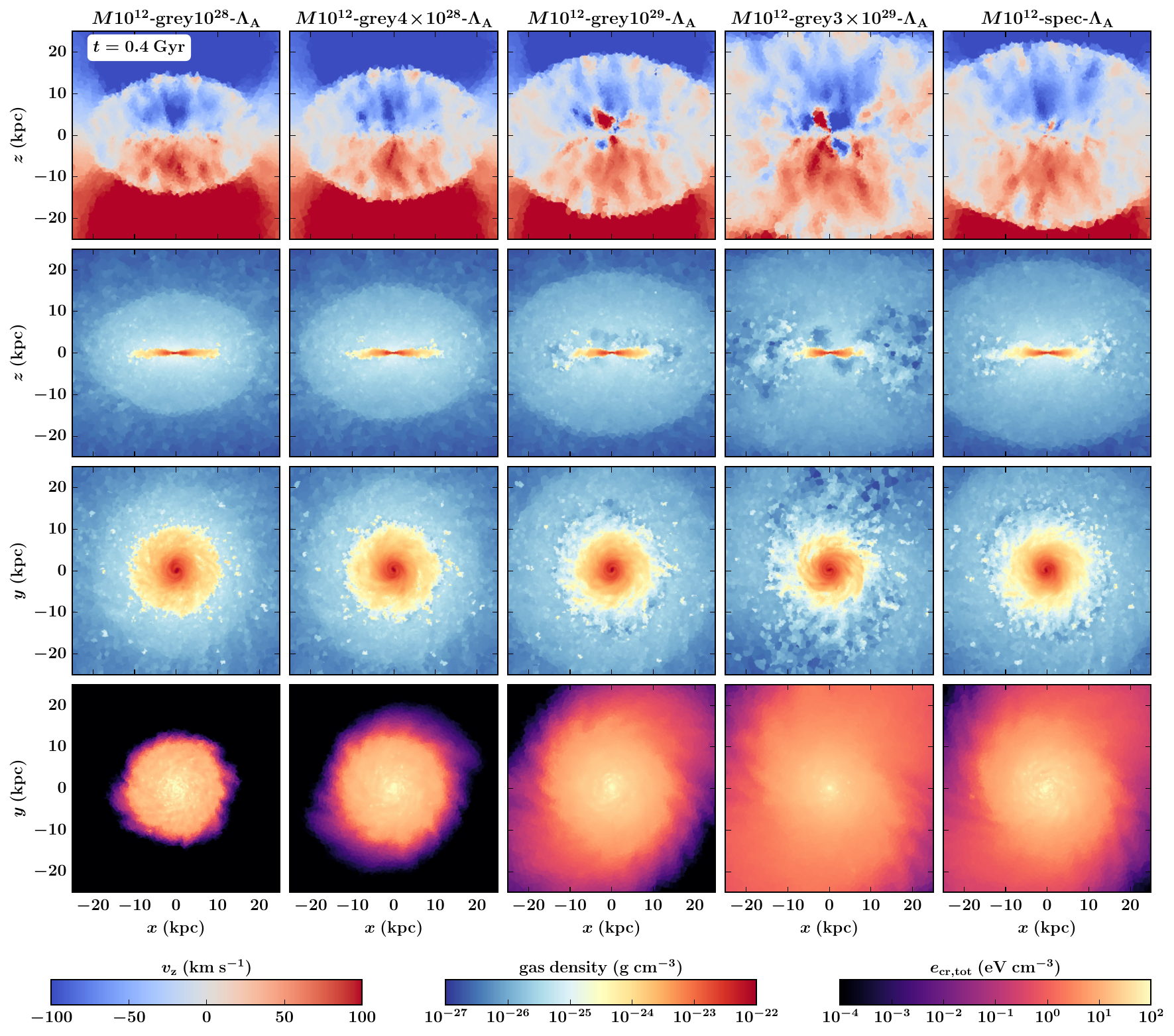}
    \caption{Onset of the outflow. The four left-hand columns depict the grey models with increasing diffusion coefficient ``$M10^{12}\text{-}\mathrm{grey}10^{28}\text{-}\Lambda_\mathrm{A}$'', ``$M10^{12}\text{-}\mathrm{grey}4\!\times\!10^{28}\text{-}\Lambda_\mathrm{A}$'', ``$M10^{12}\text{-}\mathrm{grey}10^{29}\text{-}\Lambda_\mathrm{A}$'', ``$M10^{12}\text{-}\mathrm{grey}3\!\times\!10^{29}\text{-}\Lambda_\mathrm{A}$'' while the right-hand column shows the spectral model ``$M10^{12}\text{-}\mathrm{spec}\text{-}\Lambda_\mathrm{A}$''. The top two rows show the edge-on velocity and density structure while the two bottom rows display the face-on density and total CR energy density, respectively. With increasing diffusion coefficient, the outflow front expands faster. At this early stage the spectral model behaves similarly as the grey simulation with $D=10^{29}\,\diffunit$. The main structure of the disc is very similar for most cases; only the grey model with $D=3\times10^{29}\,\diffunit$ forms a noticeably smaller disk.}
    \label{fig:maps-onset-of-outflow}
\end{minipage}
\end{figure*}

The dynamical evolution proceeds very similarly for all setups\footnote{We note that the strong difference between the grey and spectral model in \citetalias{GirichidisEtAl2022} is due to erroneously not accounting for CR cooling in the spectral model, see erratum.}. The initial gas cloud rapidly starts cooling inside out, collapses and forms stars in close proximity to the centre of the halo. The strength of this starburst phase depends on the mass of the system, with higher star formation rates for larger masses. The initial rotation of the gas cloud and the conservation of specific angular momentum during the collapse causes the formation of a disc inside out. This first phase of the gravitational collapse is dominated by adiabatic compression of the gas \citep{PfrommerEtAl2022}. The impact of the CRs is limited during this phase since they are only injected during the initial burst of star formation and need time to build up relevant energy densities.

Before investigating the details we confirm that the star formation rate in the simulations scales with gas surface density as expected (Fig.~\ref{fig:KS-relation}). Included are all models at each simulated halo mass over the full time evolution. The differences between the individual models are small (see below). Overplotted is the classical Kennicutt-Schmidt relation \citep{Schmidt1959,KennicuttSchmidt1998}. We identify the newly formed stars in patches of $(2\,\mathrm{kpc})^2$ in the $xy$-plane and compute the corresponding column density along the $z$-direction in the range of $\pm5\,\mathrm{kpc}$ around the midplane. The modelled systems with different halo masses mainly occupy different regions in the plot. At low surface densities, we note a significant deviation from the classical relation towards lower star formation rate surface densities, which is broadly in agreement with observations \citep[e.g.,][]{BigielEtAl2008,LeroyEtAl2008}

Figure~\ref{fig:sfr-outflow-time-evol} (left-hand panels) shows the time evolution of the star formation rate for all models. From top to bottom we increase the halo mass. For the early starburst phase we only find strong differences in the star formation rate between the grey and spectral CR models for the $M=10^{10}\,\Msol$ halo. The low-mass systems are less strongly dominated by gravitational collapse, so that details in the injection rate, the location and the cooling can alter the temporal evolution. Until a time of $t\sim0.5\,\Gyr$, the spectral model behaves more similar to the ``$M10^{10}\text{-}\mathrm{grey}4\times10^{28}\text{-}\Lambda_\mathrm{A}$'' run, whereas afterwards it mimics a decline that is comparable in shape to the ``$M10^{10}\text{-}\mathrm{grey}10^{28}\text{-}\Lambda_\mathrm{A}$'' setup, however, with an offset in amplitude so that it produces the largest star formation rate over the simulated time. At late times we also note higher rates for the grey run that includes Alfv\'{e}n cooling ``$M10^{10}\text{-}\mathrm{grey}10^{28}\text{-}\Lambda_\mathrm{A}$'' compared to ``$M10^{10}\text{-}\mathrm{grey}10^{28}\text{-}\Lambda_\mathrm{A}=0$'', which stems from the more efficient CR cooling and the resulting weaker CR support against gravitational collapse. For the $M=10^{11}\,\Msol$ halos, the differences in the starburst phase are negligible. At a later stage, the star formation rates in the spectral models do not decline as strongly in comparison to the grey counterparts. Towards the end of the simulation the missing Alfv\'{e}n cooling (dashed lines) again suppresses star formation in comparison to the counterpart runs that include $\Lambda_\mathrm{A}$. For the two largest halo masses, the impact of the gravitationally driven collapse of the halo gas overlies the weaker dynamical differences between the different CR models. Hence, the temporal evolution of the star formation rate does not indicate strong systematic differences. The systematic reduction of the star formation rate due to missing Alfv\'{e}n cooling as observed in the smaller mass haloes is subdominant in the temporal scatter of the largest halo. At late times the star formation rate in the grey run with $D=10^{29}\,\diffunit$ declines noticeably faster than in all other runs.

We quantify the outflows  as a function of time in Fig.~\ref{fig:sfr-outflow-time-evol} (right-hand panels). The mass flux is measured at a height of $|z|=10\,\kpc$ through a circle with radius $r=15\,\kpc$. The initial collapse of the gas is imprinted in the early phase, i.e.\ for $t\lesssim0.5\,\Gyr$, where all simulations show negative outflow rates. In the case of the low halo mass ($10^{10}\,\Msol$), the stellar feedback reverts the flow to an effective outflow at $t\approx0.5\,\Gyr$. The strongest outflows for these dwarf galaxies are driven by model ``$M10^{10}\text{-}\mathrm{grey}4\times10^{28}\text{-}\Lambda_\mathrm{A}$'' whilst the grey model with a smaller diffusion coefficient, ``$M10^{10}\text{-}\mathrm{grey}10^{28}\text{-}\Lambda_\mathrm{A}$'', and the spectral model ``$M10^{10}\text{-}\mathrm{spec}\text{-}\Lambda_\mathrm{A}$'' are almost indistinguishable for the entire simulation time. Stronger mass outflow rates are triggered in the $10^{11}\,\Msol$ models. Again, the grey model ``$M10^{11}\text{-}\mathrm{grey}4\times10^{28}\text{-}\Lambda_\mathrm{A}$'' drives the strongest outflows. The inclusion of Alfv\'{e}n cooling does not alter the outflow properties. The net outflow remains positive for most of the simulated time, but declines to irrelevant net values towards the end of the simulation. For a halo mass of $3\times10^{11}\,\Msol$ all models are able to lift gas above the measurement height of $10\,\kpc$ during the initial starburst. Here, the grey models with $D=4\times10^{28}$, $10^{29}$, and $3\times10^{29}\,\diffunit$ drive stronger peak outflows. In the case of the spectral model ($M3\times10^{11}\text{-}\mathrm{spec}\text{-}\Lambda_\mathrm{A}$) this short period is followed by weak infall for about $500\,\Myr$. Only after $t\gtrsim1.5\,\Gyr$, the flow can be reverted to a net outflow. For the highest masses we do not find net outflows for most of the simulated time. Only the grey model with the highest diffusion coefficient ($M10^{12}\text{-}\mathrm{grey}3\times10^{29}\text{-}\Lambda_\mathrm{A}$) creates a net outflow for a short amount of time after the peak in star formation ($\sim1\,\Gyr$) and for a more evolved stage later on ($t>2\,\Gyr$).

An equally important quantity is the mass loading factor $\eta(t)=\dot{M}_\mathrm{out}(t)/\dot{M}_\mathrm{SFR}(t)$. To compute $\eta$, we evaluate outflow and star formation rates at the same time despite the fact that the effective stellar feedback that leads to the outflows at a given height has taken place earlier. With typical outflow velocities exceeding $50\,\kmpersec$ and a measurement height for the outflows at $|z|=10\,\kpc$, this results in a delay of $\lesssim200\,\Myr$, which is significantly smaller than the simulation time. To avoid time scale problems we average the mass-loading factor between $t=0.5$ and $t=2.5\,\Gyr$, which is shown in Fig.~\ref{fig:mass-loading-factor} as a function of halo mass. We apply a small horizontal shift for better readability. The largest values for $\eta$ are measured for the smallest halo and $\eta$ declines to negative values when increasing the system mass from $3\times10^{11}$ to $10^{12}\,\Msol$. The variations between the models span a factor of three to four. For the grey models we note a systematic trend. With increasing diffusion coefficient $\eta$ first increases and then decreases at a fixed halo mass. Most efficient in driving outflows are the models with $D=4\times10^{28}$ and $10^{29}\,\diffunit$. An exception is the most massive halo, where the grey model with $D=3\times10^{29}\,\diffunit$ has the largest $\eta$. The spectral model and the grey counterpart with a small diffusion coefficient ($D=10^{28}\,\diffunit$) are comparable. Switching off Alfv\'{e}n cooling noticeably changes the CR energy content in the centre of the galaxy, where most stars form and where the outflows are primarily launched. As a consequence, the two models with $\Lambda_\mathrm{A}=0$ show enhanced mass loading factors. The models with a halo mass of $3\times10^{11}\,\Msol$ indicate a similar trend, however, with smaller differences while there is no difference in the most massive halo with $10^{12}\,\Msol$.

The formation of an accretion shock that travels from the forming disc into the halo and the accompanying onset of an outflow depends noticeably on the effective diffusion coefficient. We illustrate this behaviour in Fig.~\ref{fig:maps-onset-of-outflow} for a halo mass of $10^{12}\,\Msol$. The simulations with a halo mass of $3\times10^{11}\,\Msol$ show a very similar feature; for the two small halo masses the differences between the grey and spectral runs concerning the accretion shock are very minor, which is a result of the advection dominated CR transport, see below. From top to bottom the rows depict edge-on views of the vertical velocity and the gas density, followed by face-on views of the gas density and the total CR energy density. The four left-hand columns depict the grey models with different diffusion coefficients, and the right-hand column shows the spectral model. In the latter case, the high-energy CRs with diffusion coefficients exceeding most grey counterparts result in faster transport of CR energy into the halo. This triggers a faster expansion of the accretion shock and a faster onset of the outflow. The effective diffusion coefficient of the spectral model at this early stage of the hydrodynamical evolution is comparable to the diffusion in the grey model with $D=10^{29}\,\diffunit$ (see next section). We therefore note similarities between all quantities. The faster transfer of CR energy into the outskirts of the galaxy is nicely illustrated in the bottom row of Fig.~\ref{fig:maps-onset-of-outflow}. We note that the overall shape of the galactic disc does not depend on the CR model except for the grey model with the highest diffusion coefficient, which forms a smaller disk and noticeably larger perturbations in the density structure.

\section{Effective diffusion coefficients}
\label{sec:diff-coeff}

\begin{figure*}
\begin{minipage}{\textwidth}
    \centering
    \includegraphics[width=0.95\textwidth]{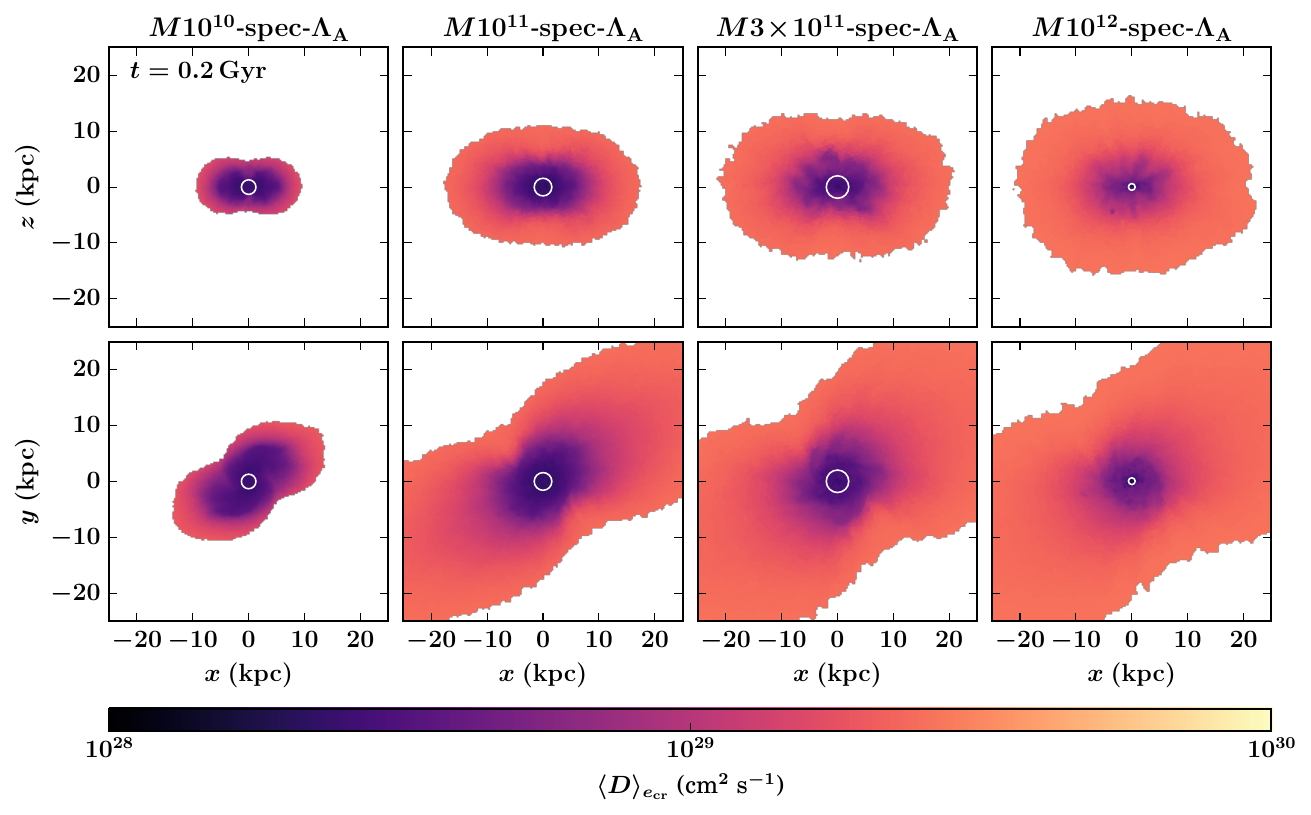}\\
    \includegraphics[width=0.95\textwidth]{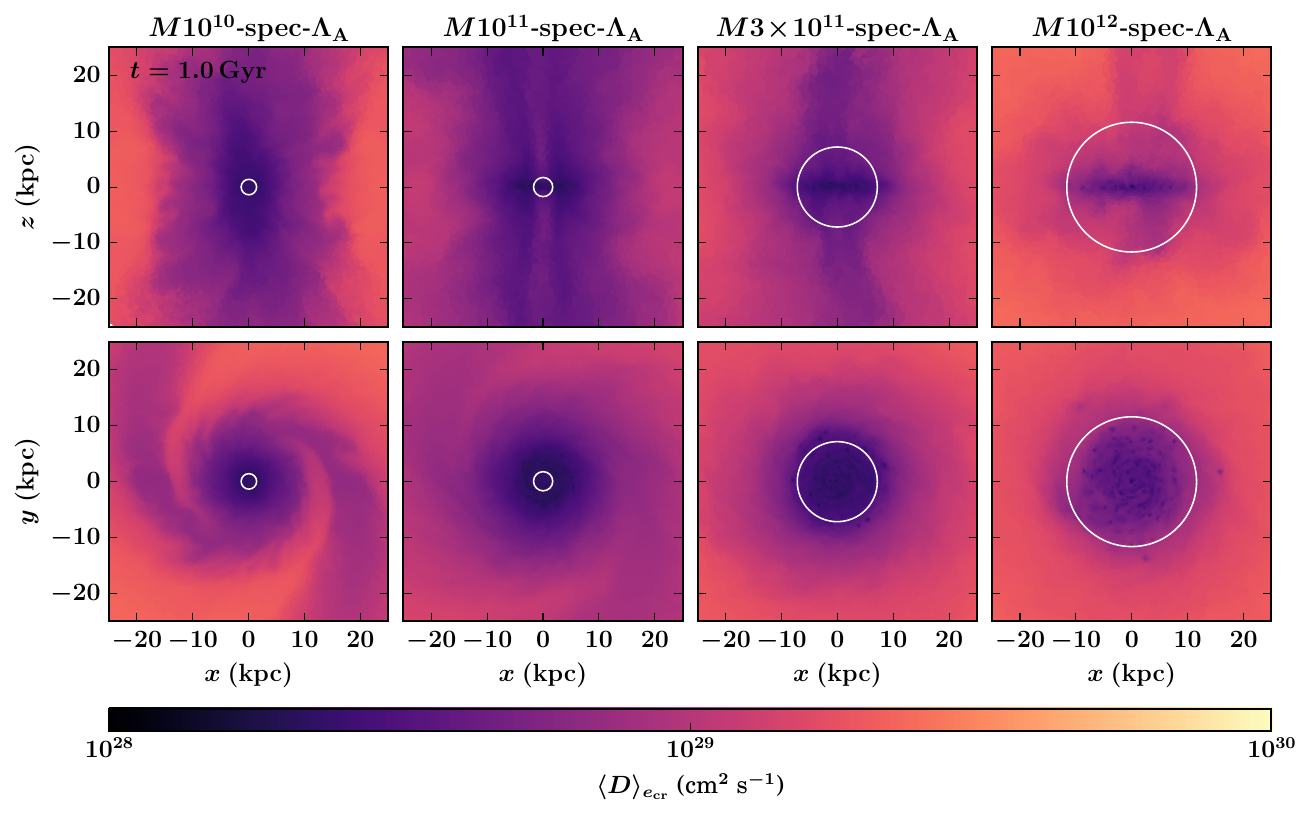}
    \caption{Effective CR energy weighted diffusion coefficient at $t=0.2\,\Gyr$ (top) and $t=1\,\Gyr$ (bottom) for models with $D_0=10^{28}\,\diffunit$. From left to right we show cuts through the middle of the box for halo masses of $10^{10}$, $10^{11}$, $3\times10^{11}$, and $10^{12}\,\Msol$. We exclude regions that are not yet populated by CRs (white areas). The circles indicate the radius of 75 per cent enclosed stellar mass. At early times we find small effective diffusivities ($\langle D\rangle_{e_\mathrm{cr}}\sim10^{28}\,\diffunit$) close to the centre, where new CRs are injected. At larger distances from the centre the spectra are more dominated by high energy CRs, which increases $\langle D\rangle_{e_\mathrm{cr}}$ to values of $\sim3\times10^{29}\,\diffunit$. At late times we note a strong difference between the regions that are dominated by advective transport with low-energy CRs dominating the spectrum and regions of diffusive transport with higher diffusivities.}
    \label{fig:diff-coeff-effective}
\end{minipage}
\end{figure*}

\begin{figure*}
\begin{minipage}{\textwidth}
    \centering
    \includegraphics[width=\textwidth]{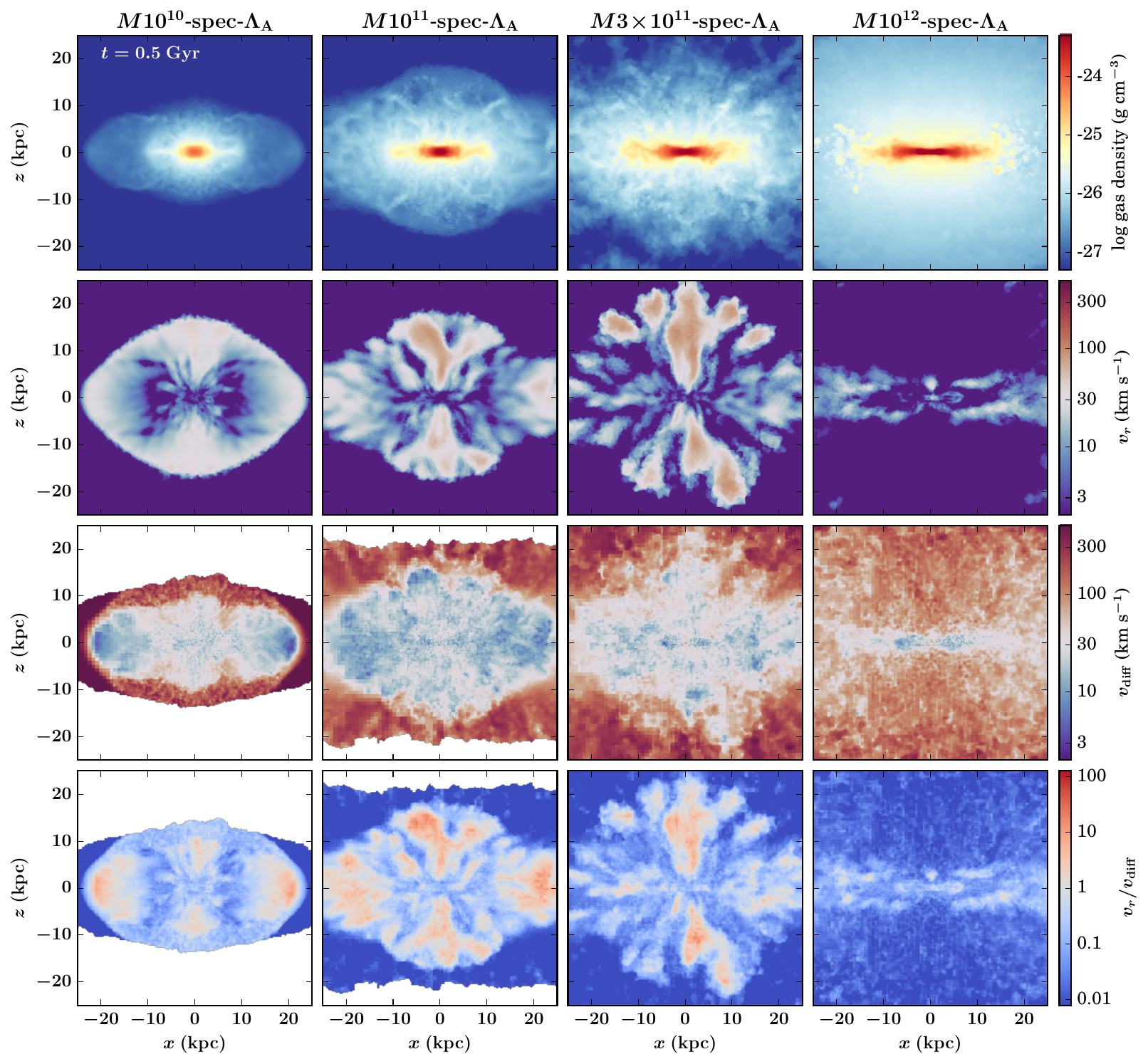}
    \caption{Comparison of the advective and diffusive CR transport speeds in the spectral CR models at $t=0.5\,\Gyr$. From left to right we increase the halo mass. From top to bottom we show edge-on maps of the gas density, the radially outward pointing advection velocity ($v_r$), the diffusion speed ($v_\mathrm{diff}$) as well as their ratio $v_r/v_\mathrm{diff}$. The low mass halo develops a smooth coherent advective outflow. The two medium mass systems show patches of strong advective motions. These features are very weak in the infall dominated high-mass halo. The ratio of the two speeds indicates that advective velocities are more dominant in low-mass systems, whereas high-mass systems are dominated by CR diffusion, see also Fig.~\ref{fig:ratio-transport-speeds-integral}.}
    \label{fig:ratio-transport-speeds-maps}
\end{minipage}
\end{figure*}

For the spectral models, we can measure the effective diffusion coefficient with which the CR energy is transported. We compute the CR energy weighted value
\begin{align}
    \langle D\rangle_{e_\mathrm{cr}} = \frac{1}{e_\mathrm{cr}}\int_{e_\mathrm{min}}^{e_\mathrm{max}} \frac{\partial e_\mathrm{cr}(p)}{\partial p}\,D(p)\dd p,
\end{align}
where  $e_\mathrm{cr}$ is the total CR energy. In our discretised spectrum this average corresponds to a sum over the spectral bins, $i$:
\begin{align}
\label{eq:D-effective}
    \langle D\rangle_{e_\mathrm{cr}} = \frac{1}{e_\mathrm{cr}}\sum_i e_{\mathrm{cr},i}(p)\,D(p_i).
\end{align}
The energy weighted diffusion coefficients are shown in Fig.~\ref{fig:diff-coeff-effective} at two different times for different halo masses. The upper panels depict an early evolutionary stage at approximately the peak of star formation ($t=0.2\,\Gyr$) while the bottom panels show a later stage at $t=1\,\Gyr$. For each time we show the edge-on view in the upper row and the face-on view in the lower one. In all cases we use the fiducial spectral model with $D_0=10^{28}\,\diffunit$, in which the diffusion coefficient scales as $D(p)\propto p^\delta$ with $\delta=0.3$. We exclude regions with a total CR energy density ($e_\mathrm{cr}$) below $10^{-5}\,\mathrm{eV\,cm^{-3}}$ (white patches) because they correspond to pristine conditions in our initial conditions. The white circle indicates the radius enclosing 75 per cent of the stellar mass to highlight the primary regions of CR injection. At the peak of the star formation rate, the central parts of the galaxies are dominated by spectra that closely correspond to the injection spectra with an abundant low-energy CR component. Hence, the energy weighted diffusion coefficient is small with a value $\sim10^{28}\,\diffunit$. The regions between the star forming sites and the accretion shock shows values of $\langle D\rangle_{e_\mathrm{cr}}\gtrsim10^{29}\,\diffunit$. The relative fraction of the volume with low values of $\langle D\rangle_{e_\mathrm{cr}}$ is much larger in the low-mass systems in comparison to the higher mass halos so that in model ``$M10^{10}\text{-}\mathrm{spec}\text{-}\Lambda_\mathrm{A}$'' most of the volume is dominated by diffusion coefficients $\langle D\rangle_{e_\mathrm{cr}}\sim10^{28}\,\diffunit$. Contrary, in run ``$M10^{12}\text{-}\mathrm{spec}\text{-}\Lambda_\mathrm{A}$'' the region with $\langle D\rangle_{e_\mathrm{cr}}\sim10^{28}\,\diffunit$ is concentrated very closely to the centre.

Low values of $\langle D\rangle_{e_\mathrm{cr}}$ indicate more abundant low-energy CRs compared to their high-energy counterparts. The existence of large volumes with small effective CR diffusion coefficients beyond the injection sites (Fig.~\ref{fig:diff-coeff-effective}) can have several reasons. First, a likely possibility is advection of CRs, which preserves their spectral shape. A second possibility is adiabatic cooling, which shifts the spectrum to lower momenta and thus gives the low-energy part a larger weight. The large volume with overall lower $\langle D\rangle_{e_\mathrm{cr}}$ in low-mass halos could also be due to a different calorimetric fraction across the different halo masses. If CRs cool less efficiently at the smaller column densities in dwarf galaxies, low-energy CRs survive for a longer time while high-energy CRs diffuse to larger distances. We note that the three possibilities are connected in a complicated way. A different effective transport is likely to result in different cooling efficiencies. Adiabatic compression or expansion shifts the spectrum and thus changes both the cooling efficiencies as well as the effective transport. We will address the details of the spectral CR shape and the calorimetry of individual energies in a follow-up paper and focus here on the effect of the different transport mechanisms. High-energy CRs can faster escape the star forming regions, which leads to spectra that are more dominated by high-energy CRs (see Girichidis et al. in prep.). In contrast, low-energy CRs need much longer to escape their acceleration sites. If they do not cool as fast as the high-energy ones diffuse away, the spectra peak at higher energies and thus increase the effective diffusivity \citep[see also][]{WerhahnEtAl2023}.

At later times ($t=1\,\Gyr$, Fig.~\ref{fig:diff-coeff-effective}, bottom panels) the values for $\langle D\rangle_{e_\mathrm{cr}}$ differ more strongly between the models, which directly reflects their dynamical state. The simulation ``$M10^{10}\text{-}\mathrm{spec}\text{-}\Lambda_\mathrm{A}$'' drives weak outflows. In spatial regions where CRs are mainly advected and where they have developed a steady state, the CR spectra are dominated by low- to medium-energy CRs, which results in effective diffusivities of $\langle D\rangle_{e_\mathrm{cr}}\sim10^{28}\,\diffunit$. This is particularly prominent in the strong outflow of model ``$M10^{11}\text{-}\mathrm{spec}\text{-}\Lambda_\mathrm{A}$''. The larger gravitational attraction in the halo masses $3\times10^{11}$ and $10^{12}\,\Msol$ results in weaker outflows, and therefore a smaller fraction of advected CR spectra.

\section{CR advection vs. diffusion dominated systems}
\label{sec:adv-vs-diff}

\begin{figure}
    \centering
    \includegraphics[width=8cm]{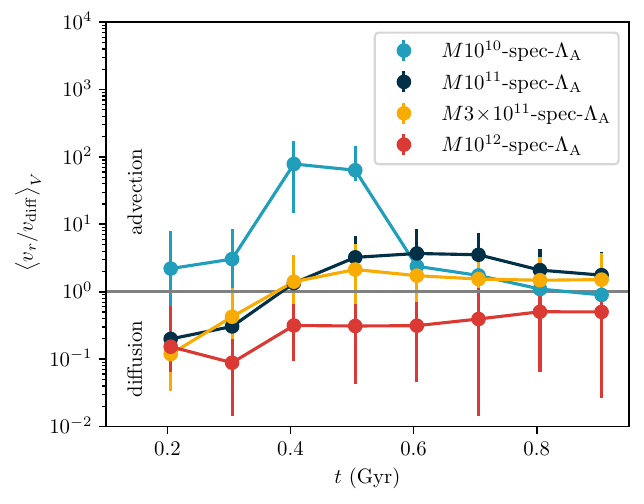}
    \caption{Volume averaged ratio  $\langle v_r/v_\mathrm{diff}\rangle_V$ within a cylinder of radius $r=25\,\kpc$ and height $\pm12.5\,\kpc$ centred on the galactic centre. From low to high masses there is a transition from advectively to diffusively dominated CR transport. The two intermediate mass halos start with diffusive CR transport before strong continuous CR-driven outflows develop so that CRs are primarily transported via advection.}
    \label{fig:ratio-transport-speeds-integral}
\end{figure}

\begin{figure*}
\begin{minipage}{\textwidth}
    \centering
    \includegraphics[width=\textwidth]{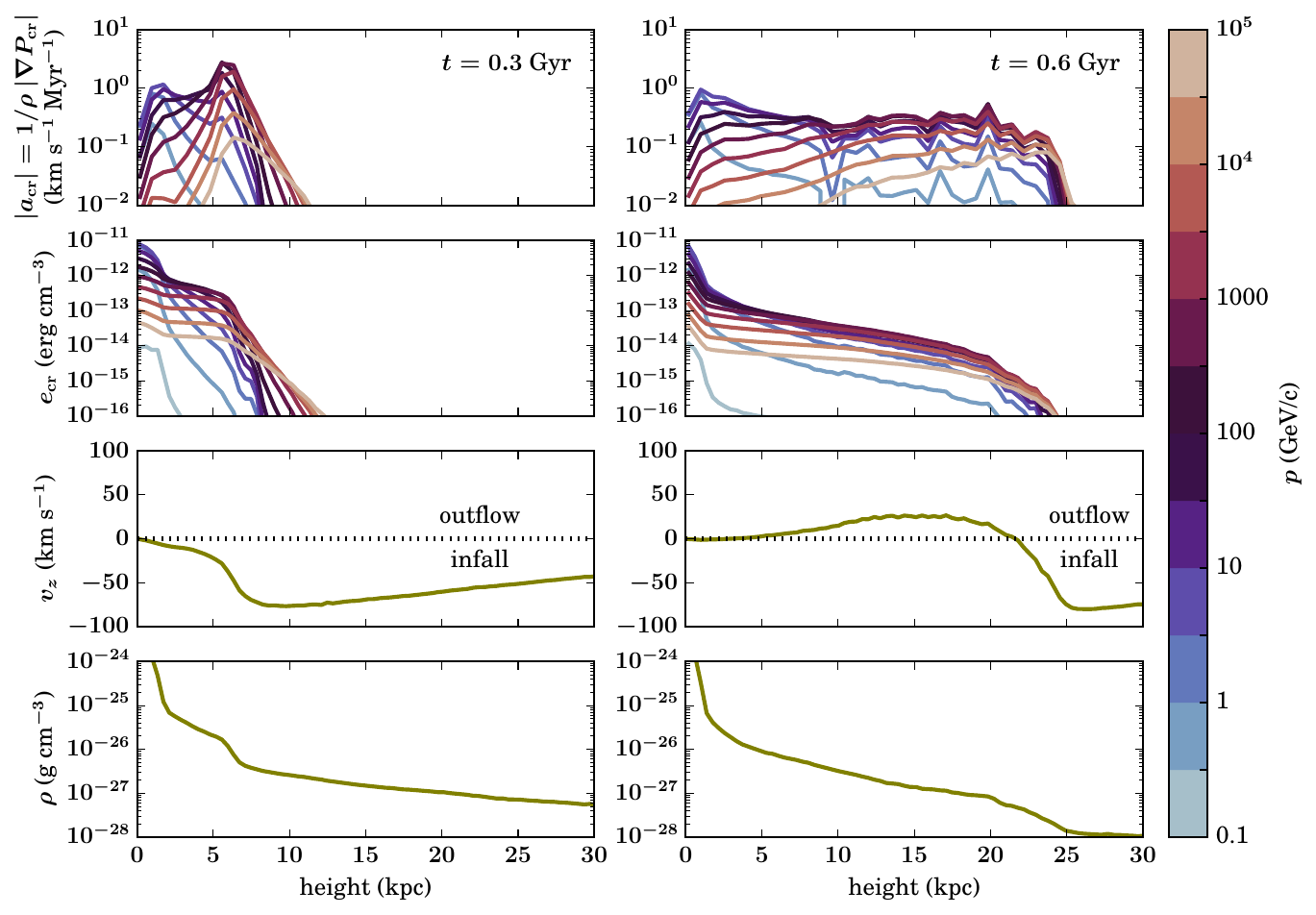}
    \caption{Vertical profiles averaged over a cylinder with radius $r=15\,\kpc$ and height $\pm30\,\kpc$ positioned on the centre of the galaxy at $t=0.3\,\Gyr$ (left) and $t=0.6\,\Gyr$ (right) for simulation ``$M10^{11}\text{-}\mathrm{spec}\text{-}\Lambda_\mathrm{A}$''. From top to bottom we show the acceleration of the gas owing to CRs, their energy density, the gas velocity as well as the gas density. Left: The early slowing-down of the infall velocity at $z\sim6-7\,\kpc$ coincides with a peak of the gas acceleration due to CRs with a particle momentum of $\sim100\,\GeVc$. Right: The infalling gas has been converted to a coherent outflow up to $z\approx22\,\kpc$. The distribution of CRs extends to a similar height ($z\sim25\,\kpc$), which is a signature of advection dominated CR transport. This is also seen as a small enhancement in the density profile. The outflow front is driven by intermediate- to high-energy CRs, whereas the acceleration of the gas at lower heights is dominated by CRs with a particle momentum of $p\sim10\,\GeVc$.}
    \label{fig:vertical-profiles-forces-1e11}
\end{minipage}
\end{figure*}

\begin{figure*}
\begin{minipage}{\textwidth}
    \centering
    \includegraphics[width=\textwidth]{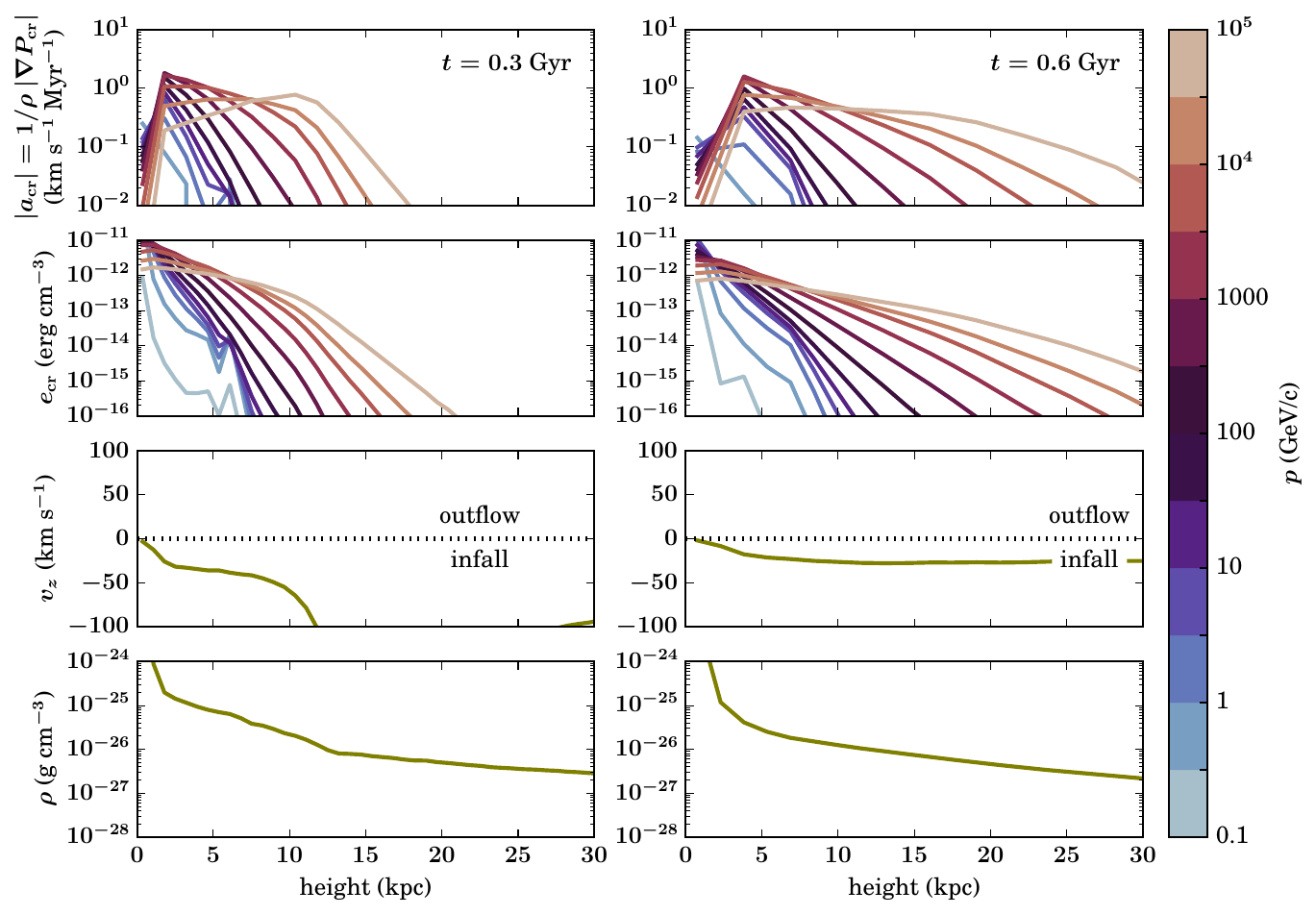}
    \caption{Same as Fig.~\ref{fig:vertical-profiles-forces-1e11} but for the more massive galaxy ($M10^{12}\text{-}\mathrm{spec}\text{-}\Lambda_\mathrm{A}$) at $t=0.3\,\mathrm{Gyr}$ (left) and $t=0.6\,\Gyr$ (right). At both times we note a significantly larger spatial spread of the different CR energies in both the CR energy density as well as the CR-driven acceleration. This large spread is attributed to energy-dependent CR diffusion and highlights the diffusion dominated transport. The infall at $t=0.3\,\mathrm{Gyr}$ is significantly slowed down but not fully stopped by CRs with momenta of $p\sim10^4\,\GeVc$ for a height $\lesssim10\,\kpc$. The overall larger central densities result in more efficient CR cooling, in particular for low-energy CRs that cannot diffuse fast enough out of the dense gas. As a result, CRs with $p\lesssim30\,\GeVc$ neither dominate the energy nor the acceleration profile.} 
    \label{fig:vertical-profiles-forces-1e12}
\end{minipage}
\end{figure*}

We investigate the relative importance of advective and diffusive CR transport in the outflow. We note that the role of advection is twofold: while CRs are advected with the flux-frozen magnetic fields and thus the outflowing gas (in addition to diffusive CR transport along the orientation of the local magnetic field), the observed galactic outflows have been driven by the CR pressure gradients. 

For this analysis we focus primarily on the early evolution during the starburst phase. In the case of advection, we only consider radially outflowing velocities, $v_r>0$, where we define
\begin{equation}
  v_r = \frac{\bvel\bcdot\vektor{r}}{|\vektor{r}|}.
\end{equation}
The diffusive speed is computed as
\begin{equation}
  v_\mathrm{diff} = - \langle D\rangle_{e_\mathrm{cr}} \frac{ \vektor{b}\bcdot\vnabla P_\mathrm{cr}}{P_\mathrm{cr}},
\end{equation}
where we use the energy weighted effective diffusion coefficient $\langle D\rangle_{e_\mathrm{cr}}$ from Eq.~\eqref{eq:D-effective}.

Figure~\ref{fig:ratio-transport-speeds-maps} illustrates these different velocities in edge-on views for our fiducial spectral CR model, with increasing halo mass from left to right. From top to bottom we show the gas density, the radially outward pointing component of the velocity $v_r$, the diffusive velocity, and the ratio of the latter two, $v_r/v_\mathrm{diff}$. In the maps of $v_\mathrm{diff}$ and $v_r/v_\mathrm{diff}$ we again exclude regions which have not been populated by injected CRs, i.e., regions with $e_\mathrm{cr}<10^{-5}\,\mathrm{eV\,cm^{-3}}$ (white areas). The maps show averages along the line of sight in a slice with a thickness of $5\,\kpc$ centred around the centre of the galaxy. For the lowest mass galaxy, the radial velocities indicate the relatively smooth dynamical structure in a coherent outflowing bubble. With increasing halo mass the perturbations in the CGM increase. The two intermediate-mass models show patches of outflowing gas. In the Milky Way-like model, the motions are dominated by infall so that we do not detect large regions of a sustained radial outflow. 

The diffusive speeds in the lowest mass halo clearly reflect the coherent radial outflow: the edge to the white areas of negligible CR energy (pristine gas of the initial conditions) approximately coincides with the outer edge of the advective bubble in the map of $v_r$. Furthermore, the diffusive speeds indicate that the outflow is dominated by advection: the locations of the edge of the visible bubble in the $v_r$ map correspond to very large diffusive speeds. Those large values arise from strong CR gradients, which are established by advecting most of the CR energy up to this edge. If diffusion were to be dominant throughout the outflow, a larger fraction of the CR energy would be able to diffuse ahead of the radially driven advective outflow, which would lower the CR gradients and the resulting diffusive speeds. This transition is clearly visible for the intermediate-mass models. The Milky Way analogue develops volume filling diffusive speeds between 30 and $100\,\kmpersec$ without a clear correlation between the advective and diffusive patterns. This overall behaviour is also reflected in the velocity ratios (bottom row in Fig.~\ref{fig:ratio-transport-speeds-maps}). The patches of regions where CR transport is primarily advective reduce with increasing halo mass.

We also look at this in a more quantitative manner by computing the volume averaged ratio  
\begin{equation}
    \Bigg\langle \frac{v_r}{v_\mathrm{diff}}\Bigg\rangle_V = V^{-1}\,\sum_i V_i \frac{v_r}{v_\mathrm{diff}},
\end{equation}
where $\sum_i$ indicates the sum over all cells $i$ with their individual volumes $V_i$, and $V=\sum_i V_i$ is the total volume. We investigate a cylinder with radius $25\,\kpc$ and vertical extent $\pm12.5\,\kpc$ centred on the galaxy so that $V=\pi 25^2 \times 25\,\kpc^3$. The time evolution is shown in Fig.~\ref{fig:ratio-transport-speeds-integral} for the same four models as in Fig.~\ref{fig:ratio-transport-speeds-maps}. In the low-mass halo, CR transport is mainly dominated by advection for most of the time shown here. Only at the end, when no net outflows are driven, patches of outflowing gas are dominated by CR diffusion. The two intermediate-mass models indicate a transition from diffusion-dominated CR transport in the beginning to a more advective CR transport at later times. This corresponds to the initial infall, which is later converted to a coherent outflow. The outflows are strong and fast enough such that advective CR speeds are more important in comparison to CR diffusion. In contrast, the Milky Way analogue is dominated by diffusive speeds over the entire simulation time.

We further investigate the contribution of the individual CR particle energies to the dynamics. We do this by analysing vertical profiles in a cylinder of radius $r=15\,\kpc$ and a height of $\pm30\,\kpc$ that is centred on the galaxy. Figures~\ref{fig:vertical-profiles-forces-1e11} and \ref{fig:vertical-profiles-forces-1e12} show the profiles for models ``$M10^{11}\text{-}\mathrm{spec}\text{-}\Lambda_\mathrm{A}$'' and ``$M10^{12}\text{-}\mathrm{spec}\text{-}\Lambda_\mathrm{A}$'', respectively. In each figure the left-hand panels show the profiles at $t=0.3\,\Gyr$ while the right-hand panels display the advancement of the outflow at $t=0.6\,\Gyr$. The top two panels show profiles of the gas acceleration driven by CRs and the CR energy density, with the color of the lines indicating the CR momentum. The lower two panels show the vertical velocity and density, respectively. All panels show volume-weighted, azimuthally averaged quantities. 

In the case of the $10^{11}\,\Msol$ halo (Fig.~\ref{fig:vertical-profiles-forces-1e11}), the central region is dominated by low-energy CRs with $p\lesssim30\,\GeVc$ at both times. At the early stage, the dominant driver of the gaseous outflows are CRs with a momentum of $p\sim100\,\GeVc$ (via the energy-dependent CR acceleration $|\vektor{a}_\mathrm{cr}|$). The peak acceleration coincides with both, an increase in velocity as well as a decrease in density. This indicates that this outflow has been launched from the centre of the galaxy. The acceleration profiles of the individual CR momenta are all spatially concentrated at around $z=6\,\kpc$ with a small spread of $\sim\pm5\,\kpc$. At later times, the inflow is reverted to an outflow, with the outflow front reaching a height of $\sim22\,\kpc$ at $t=0.6\,\Gyr$. The CR energy density and the acceleration profiles have not advanced beyond the point of the actual outflow. This emphasizes the advection dominated transport of CRs, which implies that all CR momenta reach a similar distance from the centre. At first sight there is an inconsistency between the CR transport in model ``$M10^{11}\text{-}\mathrm{spec}\text{-}\Lambda_\mathrm{A}$'' in Fig.~\ref{fig:ratio-transport-speeds-integral}, where we initially find predominantly CR diffusion, and Fig.~\ref{fig:vertical-profiles-forces-1e11}, where advection clearly dominates for both times. This apparent discrepancy originates from the different volumes under consideration. The former investigates a larger cylindrical volume, which encompasses a significant fraction of the galactic disc. The latter analysis only focuses on the outflow cone with a strong direct CR driven impact.

The situation is different in the case of simulation ``$M10^{12}\text{-}\mathrm{spec}\text{-}\Lambda_\mathrm{A}$'' (Fig.~\ref{fig:vertical-profiles-forces-1e12}). At the early time (left-hand panels), the slowing down of the infall at $z\sim10\,\kpc$ is mainly caused by CRs with high momenta ($p\gtrsim10^4\,\GeVc$). We also note that there is a larger spread in the different CR momenta, corresponding to energy-dependent diffusion, which is typical for a diffusion dominated system. At later times, the infall is however not halted and the diffusive spread of CRs of different momenta is more pronounced.

By comparing the early times for both halo masses (Figs.~\ref{fig:vertical-profiles-forces-1e11} and \ref{fig:vertical-profiles-forces-1e12}) we note very similar maximum values of the magnitude of the outward pointing acceleration of CRs, $|\vektor{a}_\mathrm{cr}|\sim2-4\,\mathrm{km\,s^{-1}\,Myr^{-2}}$. We therefore attribute the continued infall in the more massive galaxy mainly to the stronger gravitational attraction of the halo. In the shallower potential of the $10^{11}\,\Msol$ model, the same acceleration can revert the inflow to a net outflow.

\begin{figure}
    \centering
    \includegraphics[width=8cm]{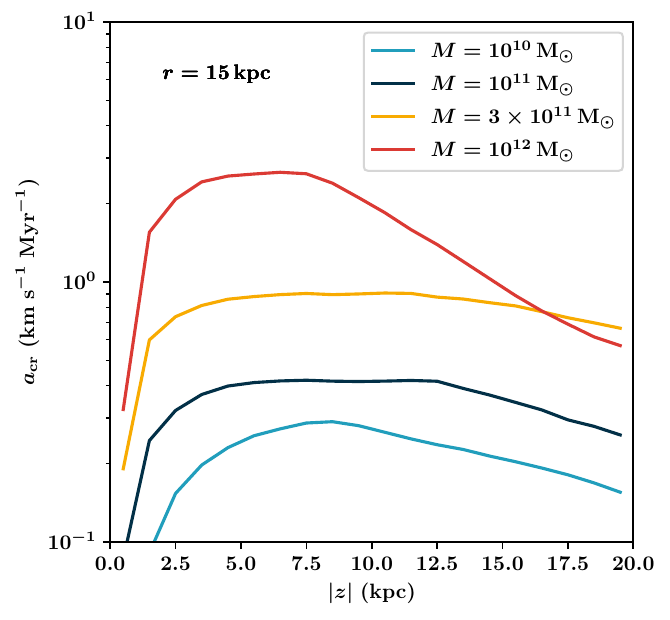}
    \caption{Comparison of time averaged ($0.5-2\,\Gyr$) vertical profiles of the CR-driven acceleration for all halo masses. We use the spectral CR model for each halo and the profiles are averaged within a cylinder of radius $r=15\,\kpc$ centred on the galaxy. The differences between the halo masses are smaller than the differences in the star formation and CR injection rate, see text.}
    \label{fig:accelerations-CR-all-halos}
\end{figure}

\begin{figure*}
    \begin{minipage}{\textwidth}
        \centering
    \includegraphics[width=0.95\textwidth]{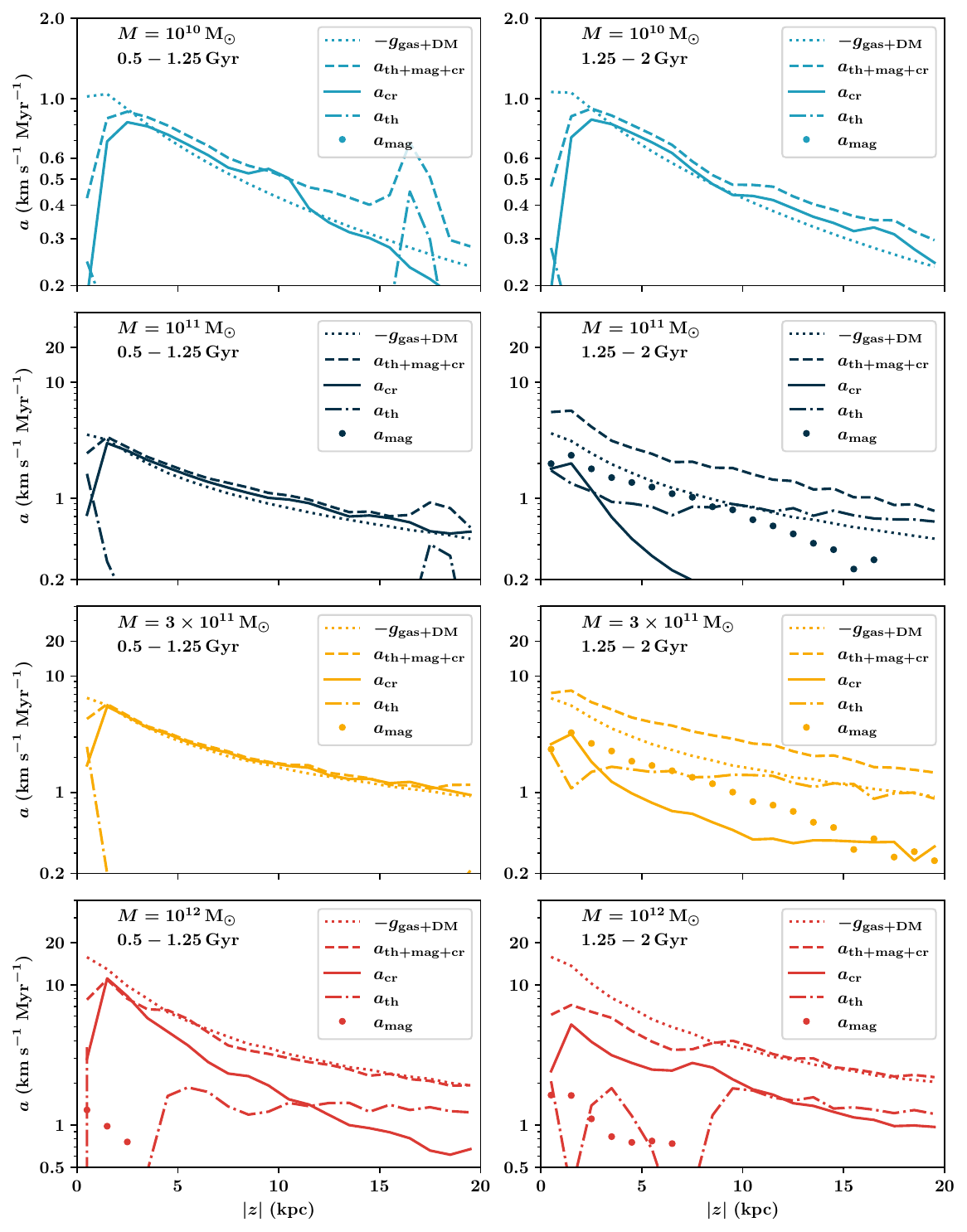}
    \caption{Time averaged vertical acceleration profiles in a cylinder with radius $r=3\,\kpc$ for the spectral CR model and all halo masses from top to bottom. The left-hand panels show the average from $0.5-1.25\,\Gyr$, the right-hand ones for $1.25-2\,\Gyr$. The low mass halo with $10^{10}\,\Msol$ is dominated by CR acceleration, which explains the strong outflows with high mass loading (see Fig.~\ref{fig:mass-loading-factor}). The two intermediate mass systems show similarities in their profiles with CRs being dominant in the beginning and supportive but not exclusively responsible for the outflow at later stages. In the most massive halo the outward pointing forces cannot compensate the gravitational attraction, i.e., no wind is launched from this galaxy.}
    \label{fig:accelerations-all-forces}
    \end{minipage}
\end{figure*}

\section{Time averaged vertical acceleration}
\label{sec:vertical-force-profiles}
We further investigate the time averaged ($0.5-2\,\mathrm{Gyr}$) vertical accelerations for the individual components that are responsible for the dynamics. Again, we focus on the spectral models for each halo mass. We would like to stress that in order to identify the long-term evolution of the galaxy and a possible continuous wind, it is important to average the accelerations over a significant fraction of the simulation time. A single acceleration profile at one point in time could also indicate a short-term fountain flow effect rather than a long-lasting outflow.

We compute the outward pointing volume-weighted time-averaged accelerations over $1.5\,\mathrm{Gyr}$ using 15 snapshots ($N_\mathrm{snap}=15$). For the CR, and thermal acceleration we compute
\begin{equation}
    a_\mathrm{th,cr} = -N_\mathrm{snap}^{-1}\sum_i^{N_\mathrm{snap}} \Bigg\langle\frac{\mathrm{sign}(z)}{\rho} \frac{\partial P_\mathrm{th,cr}}{\partial z}\Bigg\rangle_V.
\end{equation}
For the magnetic field, we need to take both magnetic pressure and tension into account,
\begin{equation}
    a_\mathrm{mag} = -N_\mathrm{snap}^{-1}\sum_i^{N_\mathrm{snap}} \Bigg\langle \frac{\mathrm{sign}(z)}{\rho}\left[\vnabla\cdot \left(\frac{\vektor{B}^2}{8\uppi}\tensor{1} - \frac{\vektor{B}\vektor{B}^\mathrm{T}}{4\uppi}\right)\right]_z\Bigg\rangle_V,
\end{equation}
where the $z$-component $[\cdot]_z$ is used for the plots. The volume average $\langle\cdot\rangle_V$ is taken over segments of cylinder with height $\Delta z=1\,\mathrm{kpc}$ and radius $r$, centred on the galaxy.

First, we quantify the similarity of the CR accelerations. We show the time averaged outward pointing total CR acceleration in Fig.~\ref{fig:accelerations-CR-all-halos} for all halo masses for the spectral CR model. The halo with $10^{12}\,\Msol$ develops a five times stronger vertical CR acceleration compared to the $10^{11}\,\Msol$ halo, whereas the peak and average star formation rates are 13 and 9 times as high. The acceleration of the $3\times10^{11}\,\Msol$ halo is twice as strong in comparison to the $10^{11}\,\Msol$ halo, which is similar to the scaling in star formation, where the peak and mean star formation rates are a factor of 2.8 higher in the larger halo. A much stronger difference is seen when comparing the $10^{10}\,\Msol$ halo with the $10^{11}\,\Msol$ counterpart. Here $a_\mathrm{cr}$ is lower by $\sim30$ per cent. The star formation rate is, however, more than an order of magnitude lower in the $10^{10}\,\Msol$ halo.


We further distinguish the time averaged accelerations between the different components in a cylinder with a radius of $3\,\kpc$ in Fig.~\ref{fig:accelerations-all-forces}. From top to bottom we increase the halo mass. In the left-hand panels we investigate the averaged accelerations from $0.5-1.25\,\Gyr$ while the right-hand counterpart shows the same quantity at a later time ($1.25-2\,\Gyr$). Shown are the total outward pointing acceleration (thermal, magnetic plus CR; dashed lines), the individual accelerations by CRs (solid lines), thermal pressure (dashed-dotted lines), and magnetic fields (dots). In addition, we plot the negative of the inward pointing gravitational attraction due to gas, stars and dark matter (dotted lines). In the case of the low-mass halo, the difference between the two averaging times does not significantly differ. Over the entire simulation time and at basically all heights the CR acceleration alone can compensate the gravitational attraction. The contribution of the other components (thermal and magnetic) is relatively small. The outflows driven from the $10^{10}\,\Msol$ halos are therefore basically purely CR driven, which is in line with the advection-dominated transport, see Section~\ref{sec:adv-vs-diff}. The more massive models with $10^{11}$ and $3\times10^{11}\,\Msol$ show a different temporal evolution. Whereas in the beginning the CRs are the main driver of the outflow at all heights, at the later stage the net acceleration of the outflow is a combination of all energy components. The most massive model with $10^{12}\,\Msol$ clearly indicates the dominance of the gravitational attraction over the different outward pointing accelerations. The temporal differences are minor with only small spatial regions showing net outward pointing accelerations that exceed the gravitational attraction. The system is therefore dominated by locally and temporally driven fountain flows. The strong gravitational attraction furthermore forces the CR transport to be dominated by diffusion rather than advection.

\section{CGM properties}
\label{sec:cgm-properties}

\begin{figure*}
\begin{minipage}{\textwidth}
    \centering
    \includegraphics[width=\textwidth]{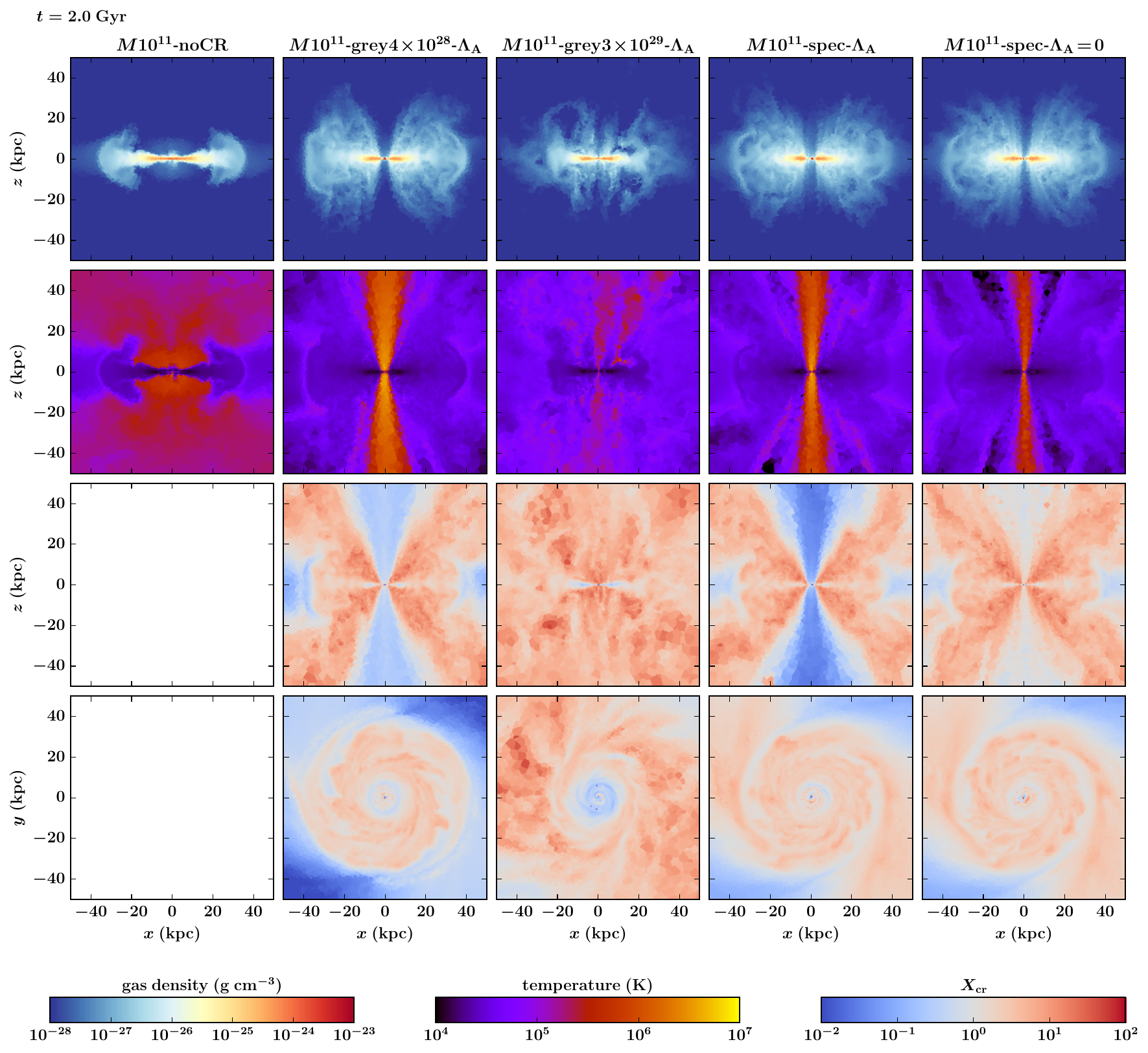}
    \caption{CGM properties for galaxies with a halo mass of $10^{11}\,\Msol$ at $t=2\,\Gyr$. From left to right we show a simulation without CRs, the grey model with large diffusivity ($M10^{11}\text{-}\mathrm{grey}4\times10^{28}\text{-}\Lambda_\mathrm{A}$) and the spectral models with and without Alfv\'{e}n cooling ($M10^{11}\text{-}\mathrm{spec}\text{-}\Lambda_\mathrm{A}$ and $M10^{11}\text{-}\mathrm{spec}\text{-}\Lambda_\mathrm{A}=0$). From top to bottom the panels depict the edge-on view of the density, the temperature and the CR-to-thermal pressure ratio $X_\mathrm{cr}$ as well as a face-on view of $X_\mathrm{cr}$. Without CRs, the CGM is filled with hot $10^{6}\,\mathrm{K}$ gas. The CR-driven outflows from the disc lead to volume filling cold gas with $X_\mathrm{cr}$ values exceeding unity. The fast outflow cones emerging from the centre of the galaxy are thermally dominated if Alfv\'{e}n cooling is included.}
    \label{fig:Xcr-maps-1e11}
\end{minipage}
\end{figure*}

\begin{figure*}
\begin{minipage}{\textwidth}
    \centering
    \includegraphics[width=\textwidth]{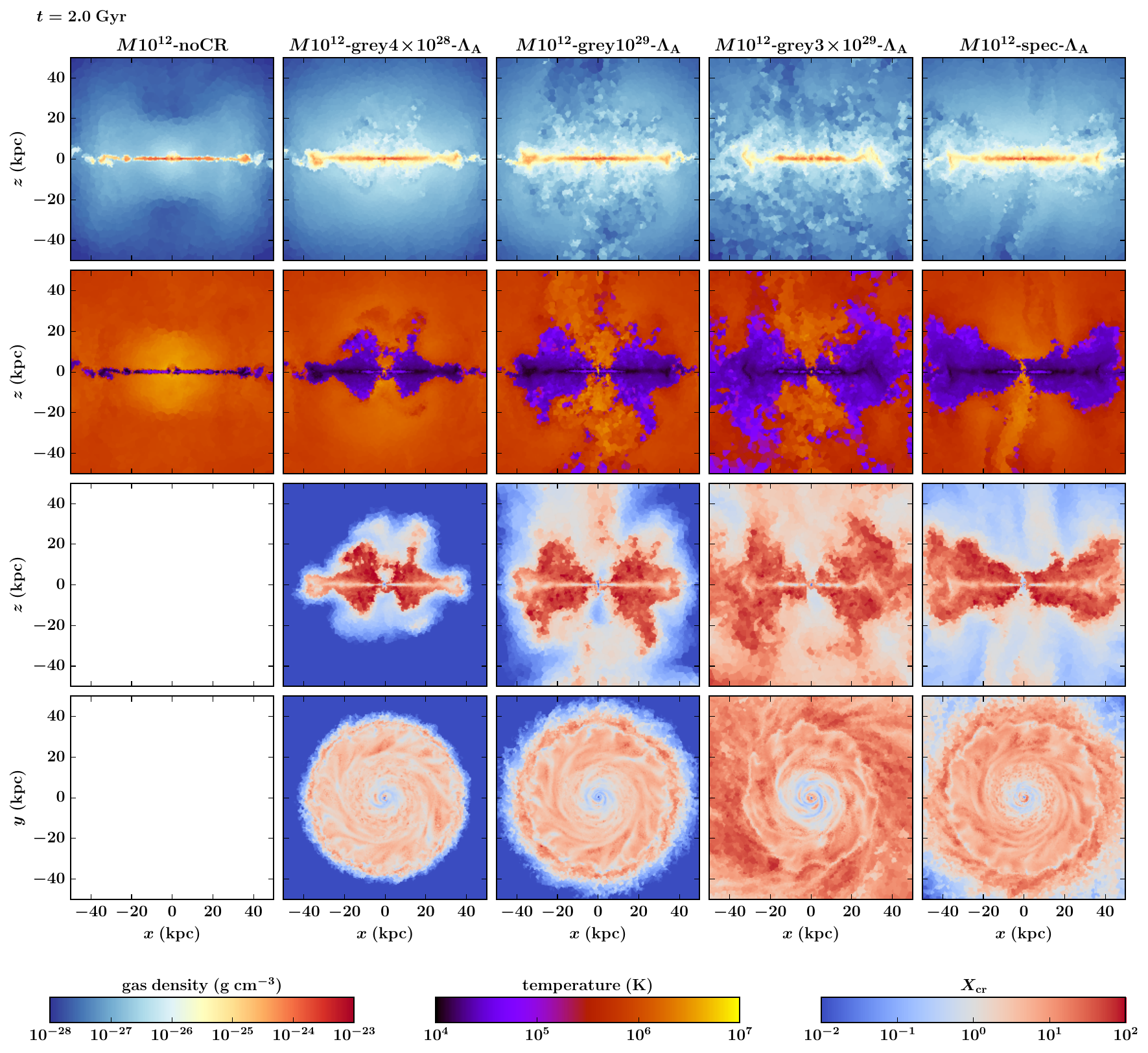}
    \caption{Same as Fig.~\ref{fig:Xcr-maps-1e11} for models with a halo mass of $10^{12}\,\Msol$ at $t=2.0\,\Gyr$. There are no significant net outflows launched for this halo mass. Nonetheless small CR-driven fountain flows cause the CGM to cool to lower temperatures in models including CRs. As a result, the cold regions are characterised by a dominating CR pressure rather than by thermal pressure.}
    \label{fig:Xcr-maps-1e12}
\end{minipage}
\end{figure*}

\begin{figure*}
\begin{minipage}{\textwidth}
    \centering
    \includegraphics[width=0.49\textwidth]{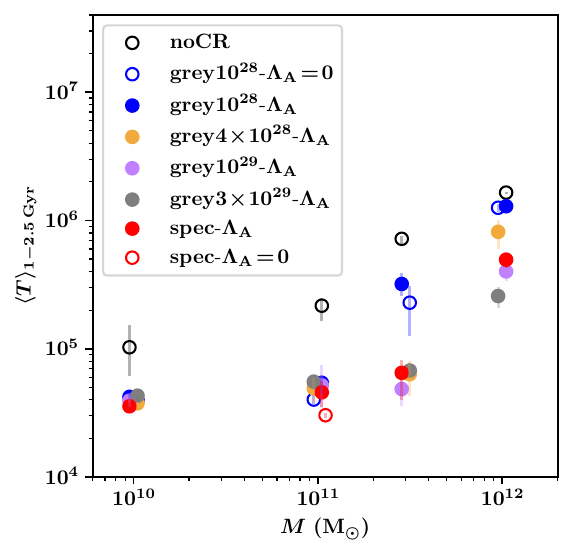}
    \includegraphics[width=0.49\textwidth]{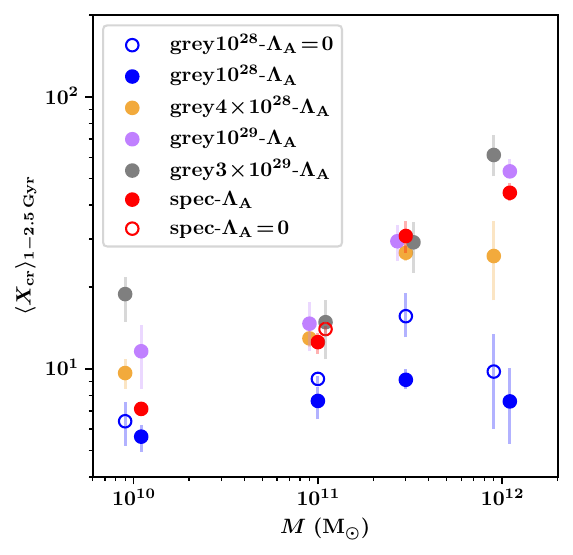}
    \caption{Time averaged ($t=1-2.5\,\Gyr$) volume weighted values of gas temperature, $T$ (left), and CR-to-thermal pressure ratio, $X_\mathrm{cr}$ (right), measured in a cylinder of radius $25\,\kpc$ and height $\pm20\,\kpc$ that is centred on the galaxy. In all cases the purely thermal runs have the highest temperature in the inspected volume. For the two lowest mass halos, the temperatures in the CR runs do not differ significantly. For the two highest mass halos, the grey runs with large diffusion coefficient and the spectral model show noticeably lower temperatures. For $X_\mathrm{cr}$ there is a general trend for all halo masses: $X_\mathrm{cr}$ increases for the grey models with increasing diffusivity. The values for the spectral models depend on their effective diffusion coefficient.}
    \label{fig:T-Xcr-time-avg}
\end{minipage}
\end{figure*}

Finally, we investigate the CGM properties of our simulated galaxies with a focus on the thermal properties and the ratio of CR-to-thermal pressure. For this comparison, we also run galaxy models without CR feedback. Figures~\ref{fig:Xcr-maps-1e11} and \ref{fig:Xcr-maps-1e12} illustrate the thermal properties for simulations with halo masses of $10^{11}\,\Msol$ and $10^{12}\,\Msol$, respectively, at $t=2\,\Gyr$. From left to right we show models with different physics parameters, the panels from top to bottom depict edge-on views of the gas density, the temperature and the CR-to-thermal pressure ratio, $X_\mathrm{cr}$, as well as a face-on view of $X_\mathrm{cr}$. For the $M=10^{11}\,\Msol$ halo, the largest systematic difference is that the run without CRs shows a much smaller vertical extent of the disc in comparison to the runs including CRs. 

The differences in the gas temperature map are even more apparent. Whereas the purely thermal simulation develops volume filling hot gas with $T\sim10^{6}\,\mathrm{K}$, the inclusion of CRs allows them to spread diffusively in the CGM, thereby providing pressure support and enabling the thermal gas to become thermally unstable and to cool down to temperatures of $T\sim10^5\,\mathrm{K}$. A noticeable exception are the outflow cones emerging from the centre of the galaxy. Here, the gas temperature exceeds $10^6\,\mathrm{K}$ for all CR models except the grey model with the highest diffusion coefficient ($D=3\times10^{29}\,\diffunit$). The difference in the CGM temperature between the grey and spectral models (``$M10^{11}\text{-}\mathrm{grey}4\times10^{28}\text{-}\Lambda_\mathrm{A}$'' vs.\ ``$M10^{11}\text{-}\mathrm{spec}\text{-}\Lambda_\mathrm{A}$'') is marginal. Switching off Alfv\'{e}n cooling increases the CR content in the outflow cone and reduces the gas temperature. In the grey model with the highest diffusion coefficient ($D=3\times10^{29}\,\diffunit$), the outflow cone is much less pronounced and the outflow region is colder. We also note morphological differences in the gas distribution. We attribute the much less pronounced outflow cone to the very fast diffusion, which in this case is larger than the energy weighted effective diffusion of the spectral model (cf. Fig.~\ref{fig:diff-coeff-effective}), and the resulting stronger CR-driven perturbations, which destroy the cone. The differences between the CR models are even stronger in the ratio $X_\mathrm{cr}$. Here, the CRs can quickly occupy the majority of the volume, which is dominated by CR pressure. In the grey CR model with $D=4\times10^{28}\,\diffunit$, the outflow cone is marginally dominated by thermal pressure. Contrary, in the spectral model, the thermal pressure in the outflow cone is an order of magnitude larger than the CR pressure. This difference is a consequence of the energy dependent diffusion. Effectively, more CR energy can escape via fast diffusion, which lowers the CR content in the outflow while the thermal structure of the outflow cone is comparable. We note that the advection dominated CR transport discussed in Section~\ref{sec:adv-vs-diff} focuses on early times and evolves to a balanced transport situation after $t\gtrsim1\,\Gyr$ (Fig~\ref{fig:ratio-transport-speeds-integral}). When Alfv\'{e}n cooling is switched off, the less efficient cooling of CRs results in equal pressures of CRs and thermal gas in the outflow cone.

For the Milky Way-mass galaxy model (Fig.~\ref{fig:Xcr-maps-1e12}), the situation is different because no significant net outflows are launched. The purely thermal simulation forms a very thin disc with no perturbations driven into the halo. This implies a hot CGM with temperatures exceeding $10^6\,\mathrm{K}$ that has been heated by the outwards propagating accretion shock. Including CRs allows for a colder environment at the disc-halo interface. This is due to CR-driven fountain flows from the galactic disc, which entrain the initially hot CGM and mix it with the colder ISM, thereby enabling thermal instability. As a result, the CRs are providing the dominant pressure support at the halo-disc interface. In this case, the CR transport speed is of importance. The grey run with a low diffusivity (``$M10^{12}\text{-}\mathrm{grey}4\times10^{28}\text{-}\Lambda_\mathrm{A}$'') drives weaker fountain flows that reach a lower altitude above the disc in comparison to the grey simulation with larger diffusion coefficients ($M10^{12}\text{-}\mathrm{grey}10^{29}\text{-}\Lambda_\mathrm{A}$ and $M10^{12}\text{-}\mathrm{grey}3\times10^{29}\text{-}\Lambda_\mathrm{A}$). The fast diffusion of the high-energy CRs in the spectral model ($M10^{12}\text{-}\mathrm{spec}\text{-}\Lambda_\mathrm{A}$) is comparable to the grey model with $D=10^{29}\,\diffunit$, which is in agreement with the energy weighted effective diffusion coefficient, see Fig.~\ref{fig:diff-coeff-effective}. The strong impact of faster diffusion on the thermal properties of the (inner) CGM is also seen in the face-on view of $X_\mathrm{cr}$ (bottom panels). Whereas the grey models with $D=4\times10^{28}$ and $10^{29}\,\diffunit$ show a sharp edge at $r\sim40\,\kpc$ beyond which no significant amount of CRs can be transported, the high-$D$ grey and the spectral run fill a larger volume with CR pressure dominated gas ($X_\mathrm{cr}>10$).

We show the time averaged values for the temperature and $X_\mathrm{cr}$ over $1.5\,\Gyr$ ($t=1-2.5\,\Gyr$) in Fig.~\ref{fig:T-Xcr-time-avg}. Both measures are volume weighted in a cylinder of radius $r=25\,\kpc$ and height $z=\pm 20\,\kpc$ around the centre of the galaxy. We define $\langle X_\mathrm{cr}\rangle_V = \langle P_\mathrm{cr}/P_\rmn{th}\rangle_V$ as the volume-weighted CR-to-thermal pressure ratio. As illustrated in Figs.~\ref{fig:Xcr-maps-1e11} and \ref{fig:Xcr-maps-1e12}, the simulations without CRs form a hotter CGM. The gas temperature increases with halo mass, which is a direct consequence of a virialisation process of the CGM driven by the expanding accretion shock. In all but the Milky Way models the difference between the non-CR and CR runs is more than a factor of two in the average temperature. In case of the two low-mass halos, the differences between the different CR physics does not alter the temperatures significantly. This is in line with the temporally strongly advective motions and the resulting outflows, where the injected CRs and the cooler gas are equally well transported throughout the inspected volume. For the two more massive halos CR diffusion plays a more important role, which is directly reflected in $\langle T\rangle$. The grey models with lower diffusivity have a weaker impact on the halo gas and lead to a higher gas temperature in comparison to the grey models with a larger diffusion coefficient and the spectral model. For $\langle X_\mathrm{cr}\rangle$ we first note that in all halos, the CR pressure dominates over the thermal pressure (in the considered volume) with a smallest value of $\langle X_\mathrm{cr}\rangle\sim5$. We observe an increasing value of $\langle X_\mathrm{cr}\rangle$ with increasing diffusivity for all halo masses, which is exactly reversed in comparison to the temperature, just as we expect if the halos (outside the fast outflow) are in approximate hydrostatic equilibrium. The lowest value of $\langle X_\mathrm{cr}\rangle$ is measured for models with $D=10^{28}\,\diffunit$. The corresponding runs without Alfv\'{e}n cooling leave more CR energy in the system, which increases $\langle X_\mathrm{cr}\rangle$.

\section{Discussion}
\label{sec:discussion}

We find that between a halo mass of $3\times10^{11}$ and $10^{12}\,\Msol$ the gravitational attraction towards the centre of mass becomes too strong for the stellar feedback to drive outflows in our setting. This result is consistent with previous studies. \citet{BoothEtAl2013} model stellar feedback including CRs in both a dwarf as well as a Milky Way galaxy and find that only the former one is able to drive outflows with mass loading factors exceeding unity. A similar limiting mass for driving sustained outflows over longer times was found in CR models by \citet{JacobEtAl2018} using an overall very similar setup like our grey models, except that they use a Hernquist \citep{Hernquist1990} instead of an NFW profile \citep{NavarroFrenkWhite1997} and isotropic CR diffusion. However, the details in the star formation rate and the mass loading factor differ. We find mass loading factors that are approximately  $\eta\sim0.5-2$ for the $10^{10}\,\Msol$ halo and $\eta\sim0.5$ for the $10^{10}\,\Msol$ model, whereas \citet{JacobEtAl2018} record $\eta\sim20-40$ and $\eta\sim2-3$ for the same halo masses mass. We note that the models that have investigated the halo mass dependence only use a simplified model for the stellar feedback. Small scale simulations of the multiphase ISM show that the details of early stellar feedback like radiation and stellar winds in combination with SN feedback and CRs can perceptibly change the effective outflows significantly \citep[e.g.][]{GattoEtAl2017, RathjenEtAl2021,SimpsonEtAl2022,KimEtAl2023}. This is particularly important in low-mass galaxies, where stellar feedback has a strong effect.

Besides the ability to drive net outflows, CRs change the thermal properties in the ISM and the outflow. \citet{GirichidisEtAl2018a} find smoother and colder gas in the presence of CRs compared to non-CR models using stratified boxes. This is consistent with full galactic models \citep{DashyanDubois2020, NunezCastineyraEtAl2023}. Apart from the thermal structure, CRs leave an imprint on the angular momentum of the outflowing gas. \citet{PeschkenEtAl2021, PeschkenEtAl2023} find that there are two different outflows phases; the warm outflows with high angular momentum and the hot outflows with low angular momentum.

We chose the simplified advection-diffusion approximation for the CR transport, which does not \emph{explicitly} model the interaction of CRs with Alfv\'{e}n modes due to the streaming instability \citep{KulsrudPearce1969, Wentzel1974,Skilling1971,Skilling1975,Shalaby2021}. The amplification of Alfv\'{e}n waves requires energy, which is provided by the CRs and causes them to lose energy. Since diffusion is energy conserving, this loss of CR energy is not included in the pure CR diffusion approach. The effect of CR losses during CR streaming can to first order be emulated in a modified diffusion approach, \citep{WienerEtAl2013, BuckEtAl2020}, which is also included in this study. This means that our effective transport speeds \emph{implicitly} depend on the effective Alfv\'{e}n wave cooling since it alters the CR energy gradients. \citet{SharmaColellaMartin2010} included CR streaming in a one-moment approach \citep[see also][]{WienerEtAl2017, HolguinEtAl2019}, in which the streaming speed needs to be regularised. However, this introduces a numerical parameter that is connected to a critical CR gradient length and causes an unphysically fast transport of CRs if they injected into an already existing distribution of background CRs \citep{ThomasPfrommer2019}.

A more fundamental approach uses two moments to compute the CR transport and the resulting energy losses. \citet{JiangOh2018,ChanEtAl2019,ArmillottaOstrikerJiang2021, ArmillottaOstrikerJiang2022} evolve the CR energy density and the momentum density with a diffusion coefficient that assumes a steady-state wave energy. More accurately, \citet{ThomasPfrommer2019, ThomasPfrommer2022} solve for the forward and backward propagating resonant Alfv\'{e}n wave energies, which allows to compute the CR diffusion coefficient more self-consistently. In these two-moment approaches the effective CR transport speeds directly depend on the efficiency of wave damping and the coupling of the CRs to the ISM. While this approach is more self-consistent, it requires an estimate of the degree of ionization of the background plasma since the efficiencies of ion-neutral and non-linear Landau damping of Alfvén waves are strong functions of the ionization degree. The resulting effective transport speed therefore sensitively depends on an accurate modeling of the different ISM phases.

We note that the numerical setup of an isolated cooling halo is ideal for systematically investigating the internal processes in the forming galaxy. However, none of the more complex environmental conditions expected in cosmological simulations are accounted for, such as the long-term evolution including accretion of gas onto the halo, development of turbulent motions and the accompanied amplification of magnetic fields in the halo \citep[e.g.][]{PakmorEtAl2020}. This emphasizes the importance of applying the spectral CR model to cosmological (zoom) setups.

\section{Conclusions}
\label{sec:conclusions}

We have performed simulations of isolated rotating cooling halos of different masses from dwarfs to Milky Way analogues with four masses ($M=10^{10}$, $10^{11}$, $3\times10^{11}$, and $10^{12}\,\Msol$). The forming galaxies use an effective model for the ISM including a stochastic star formation and SN feedback prescription, in which we only inject the CR energy for each SN directly. We inject 10 per cent of the SN energy in CRs, which we model as a dynamically coupled fluid. We distinguish between a grey CR approximation, in which only the total CR energy is evolved and a spectrally resolved counterpart, where we resolve the CR momenta ranging from the sub-relativistic ($p_\mathrm{min}=0.1\,\GeVc$) up to the ultra-relativistic regime ($p_\mathrm{max}=100\,\TeVc$). We use the advection-diffusion approximation for the CR transport with anisotropic energy dependent diffusion along the magnetic field lines. In order to account for CR losses during the transport we also apply a simplified model for CR Alfv\'{e}n wave cooling. We perform 32 simulations in total, which allow us to draw the following conclusions.

\begin{itemize}
    \item We find CR driven outflows in all but the most massive model, i.e., for all halos below the Milky Way analogue. The overall mass loading factors reduce with increasing halo mass due to the stronger gravitational attraction. In the Milky Way model ($M=10^{12}\,\Msol$) the strong initial infall of halo gas slows down over time but cannot be reverted to a net outflow. The time averaged acceleration profiles reveal that in the low-mass model CRs alone are able to drive the outflows. In the intermediate mass halos the outflows are dominantly but not exclusively driven by CRs.
    
    \item We find a clear distinction between advective and diffusive CR transport. In the galaxy with the lowest halo mass, the low gravitational impact allows the CRs to have a strong dynamical effect. Strong outflows with large mass loading factors are driven shortly after the onset of star formation. As a result, the transport of CRs is mainly advective in this model. The three more massive halos experience a stronger initial infall of gas before feedback can slow down the infall. During this initial phase the three models show a more diffusive CR transport. For the two intermediate mass halos ($M=10^{11}\,\Msol$ and $M=3\times10^{11}\,\Msol$) the onset of a delayed outflow can transform the overall diffusive to a globally advective CR transport. The outflow cone is dominated by advective transport from the very beginning. In the Milky Way model, the transport is dominated by diffusion over the entire simulated time. We clearly identify typical diffusive features where higher energy CRs reach larger distances from the injection sites despite a net infall of gas.

    \item The effective diffusion coefficients in the spectral models vary by two orders of magnitude in space and over time. In the injection regions close to the centres of the galaxies, where most stars form, the spectra are close to the injection spectra. Since we inject CRs with a power-law with $f(p)\propto p^{-4.2}$ the spectra are dominated by low energy CRs. The corresponding energy weighted diffusion coefficients are therefore smaller close to the injection sites ($D(p)\propto p^{0.3}$) with values of $\langle D\rangle_{e_\mathrm{cr}}\sim 10^{28}\,\diffunit$. At larger distances from the star forming centres, the effective diffusion coefficients depend on the dominant CR transport mode. For the models with $M=10^{10}\,\Msol$ and $M=10^{11}\,\Msol$ strong outflows lead to a predominantly advective CR transport, which keeps the shape of the spectrum and results in similarly low values of $\langle D\rangle_{e_\mathrm{cr}}\sim 10^{28}\,\diffunit$ in the outflow region. For the Milky Way model, the predominantly diffusive transport causes the spectra in outer the regions to be dominated by CRs with higher momenta. The resulting values for the effective diffusivity are then $\langle D\rangle_{e_\mathrm{cr}}\sim 2-3\times 10^{29}\,\diffunit$.
    
    \item The scaling of the CR energy density with the gas density closely follows an adiabatic scaling of $e_\mathrm{cr}\propto\rho^{4/3}$ for densities $\rho\lesssim10^{-26}\,\gpcc$. Above this density, non-adiabatic effects such as CR cooling and CR diffusion out of the dense gas flatten the scaling. Hadronic cooling scales with $\Lambda_\mathrm{hadr}\propto\rho^{7/3}$ for the density regime of adiabatic CRs. Alfv\'{e}n cooling scales with the density as predicted by simple scaling estimates with $\Lambda_\mathrm{A}\propto \rho^{11/6}$. Except for the lowest densities ($\rho\lesssim10^{-28}\,\gpcc$) the hadronic cooling dominates over the Alfv\'{e}nic counterpart, which explains the small differences in star formation and outflow rate between simulations with and without Alfv\'{e}n cooling. 
    
    \item Both star formation and outflow rates do not vary perceptibly between the different CR models. In particular, spectrally resolved CRs do not change the peak of the star formation rate or the overall temporal evolution\footnote{The strong difference to the spectral results in \citetalias{GirichidisEtAl2022} originates from the missing CR cooling.}. Minor differences are present for the lowest mass halo ($M=10^{10}\,\Msol$) due to the overall large dynamical impact of the CRs.

\end{itemize}

This demonstrates the need to account for spectral CR transport in order to provide a reliable description of the CGM as well as CR-induced non-thermal radio and gamma-ray emission \citep{WerhahnEtAl2023}.

\section*{Acknowledgements}
We thank the anonymous referee for a very careful reading of the manuscript and very valuable comments that helped improving the paper. PG, MW and CP acknowledge funding from the European Research Council under ERC-CoG grant CRAGSMAN-646955 and CP acknowledges funding from the ERC-AdG grant PICOGAL-101019746. PG also acknowledges funding from the ERC Synergy Grant ECOGAL (grant 855130).

\section*{Data availability}
The simulation data and data analysis scripts for this study will be shared upon reasonable request to the corresponding author.




\bibliographystyle{mnras}
\bibliography{girichidis.bib,astro.bib} 

\appendix
\section{Effective adiabatic index}
\label{sec:app-gamma-effective}
\begin{figure*}
\begin{minipage}{\textwidth}
\centering
\includegraphics[width=\textwidth]{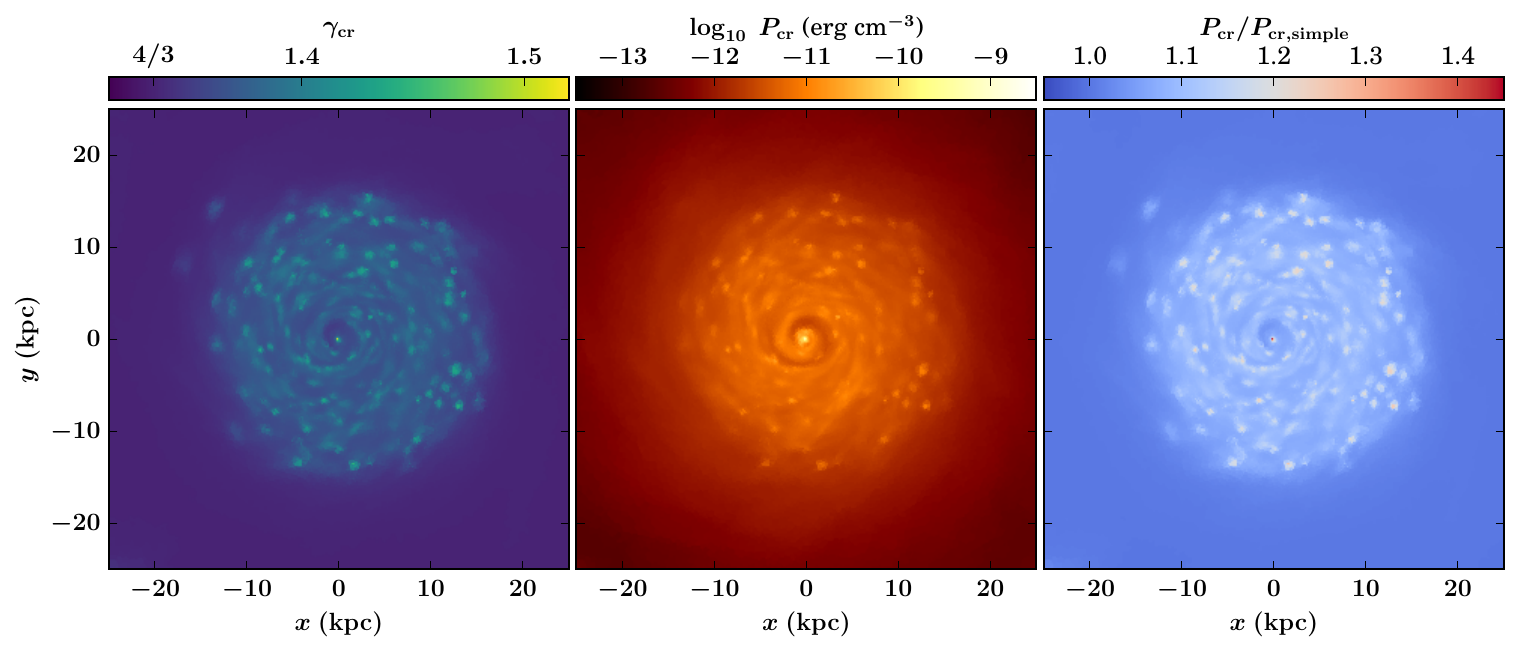}
\caption{Effective adiabatic index $\gamma_\mathrm{cr}$ (left), the total CR pressure $P_\mathrm{cr}$ (middle) and the pressure ratio $P_\mathrm{cr}/P_\mathrm{cr,simple}$ (right), where $P_\mathrm{cr,simple}$ is computed using $\gamma=4/3$. Here we show simulation ``$M10^{12}\text{-}\mathrm{spec}\text{-}\Lambda_\mathrm{A}$'' at $t=2\,\Gyr$. The other simulations show comparable or even smaller differences between the spectrally computed pressure and the simplified counterpart.}
\label{fig:effective-gamma}
\end{minipage}
\end{figure*}
Typically, the adiabatic index for CRs $\gamma_\mathrm{cr}$ is assumed to be $4/3$, which corresponds to a relativistic gas. When resolving the spectral distribution, we need to compute the effective adiabatic index based on the shape of the particle distribution function, see \citetalias{GirichidisEtAl2022}. In general the adiabatic index reads
\begin{equation}
    \gamma_{\mathrm{cr}} = \left.\frac{\dd \ln P_\mathrm{cr}}{\dd \ln\rho}\right|_S,
\end{equation}
where $S$ denotes the entropy. For our spectral model the adiabatic index can be rewritten as a weighted sum of the partial pressures in each spectral bin
\begin{equation}
    \gamma_{\mathrm{cr}} = P_{\mathrm{cr}}^{-1}\,\sum_{i=1}^{N_\mathrm{spec}}\gamma_{\mathrm{cr},i}\,P_{\mathrm{cr},i}
\end{equation}
with the total CR pressure $P_{\mathrm{cr}} = \sum_{i=1}^{N_\mathrm{spec}}P_{\mathrm{cr},i}$. We note that the adiabatic index in our case can only vary between 4/3 and 5/3, see \citetalias{GirichidisEtAl2022}. In Fig.~\ref{fig:effective-gamma} we show the effective adiabatic index (left), the pressure that is derived from the CR spectra, $P_\mathrm{cr}$ (middle) and the ratio $P_\mathrm{cr}/P_\mathrm{cr,simple}$ (right), where $P_\mathrm{cr,simple}$ is the simplified CR pressure assuming a constant adiabatic index of $\gamma=4/3$. Assuming the same CR energy, the pressure $P_\mathrm{cr} = (\gamma_\mathrm{cr} - 1) e_\mathrm{cr}$ is correspondingly larger if the distribution contains non-relativistic CRs. We measure an effective increase in pressure by approximately 20 per cent.

\section{Emulated CR streaming losses}
\label{sec:spectra}

\begin{figure*}
    \begin{minipage}{\textwidth}
    \centering
    \includegraphics[width=0.95\textwidth]{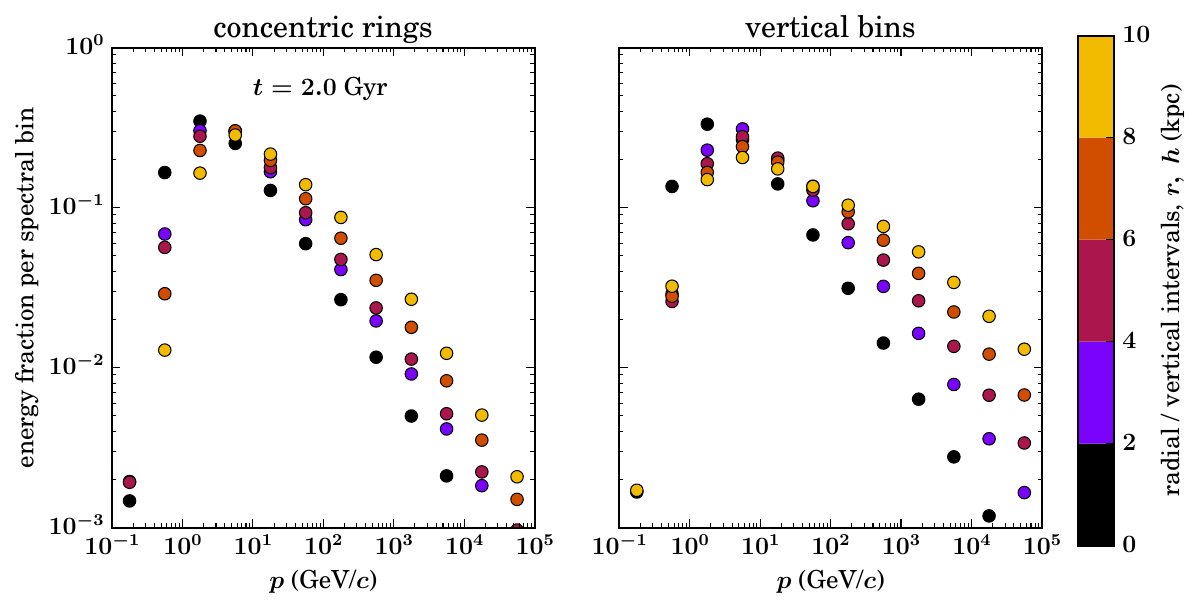}
    \caption{Relative contribution of each spectral bin to the total CR energy. We note that this quantity also measures the relative CR energy loss for the emulated CR streaming model for simulation ``$M10^{11}\text{-}\mathrm{spec}\text{-}\Lambda_\mathrm{A}$'' at $t=2\,\Gyr$. We distinguish between averaging over concentric rings (left-hand panel) and vertical heights (right-hand panel). For CR momenta with $p>1\,\TeVc$, for which CR streaming losses are likely to be unimportant, the relative fraction in total energy is negligible, so is the error due to the simplified streaming losses.}
    \label{fig:spectra-relative-energies}
    \end{minipage}
\end{figure*}

Modelling the losses due to the streaming instability should depend on the individual CR energies. In particular, the high energy CRs with momenta $p\gtrsim1\,\TeVc$ are likely not affected by the streaming losses. Our emulated streaming losses reduce the net CR energy and we simply rescale the spectrum to account for the total desired loss. This means that the effective losses in each spectral bin depend on the spectral shape, because the peaks in the spectra suffer a larger net loss compared to bins which only carry a small amount of CR energy. For most cases, the spectra peak in the range between $1-10$\,GeV\,${c^{-1}}$. The critical spectral bins (bin numbers 8-12, $p>1\,\TeVc$), only contain a small amount of total CR energy in each cell, so the error in applying a simplified cooling are expected to be small. Figure~\ref{fig:spectra-relative-energies} shows the fraction of the energy in each spectral bin for simulation ``$M10^{11}\text{-}\mathrm{spec}\text{-}\Lambda_\mathrm{A}$'' at $t=2\,\Gyr$. Shown are the radially averaged spectra (left-hand panel) as well as the vertically averaged counterparts (right-hand panel) for different concentric rings and heights, respectively.

\section{Magnetic field structure}
\label{sec:B_morphology}

\begin{figure*}
    \begin{minipage}{\textwidth}
    \centering
    \includegraphics[width=0.95\textwidth]{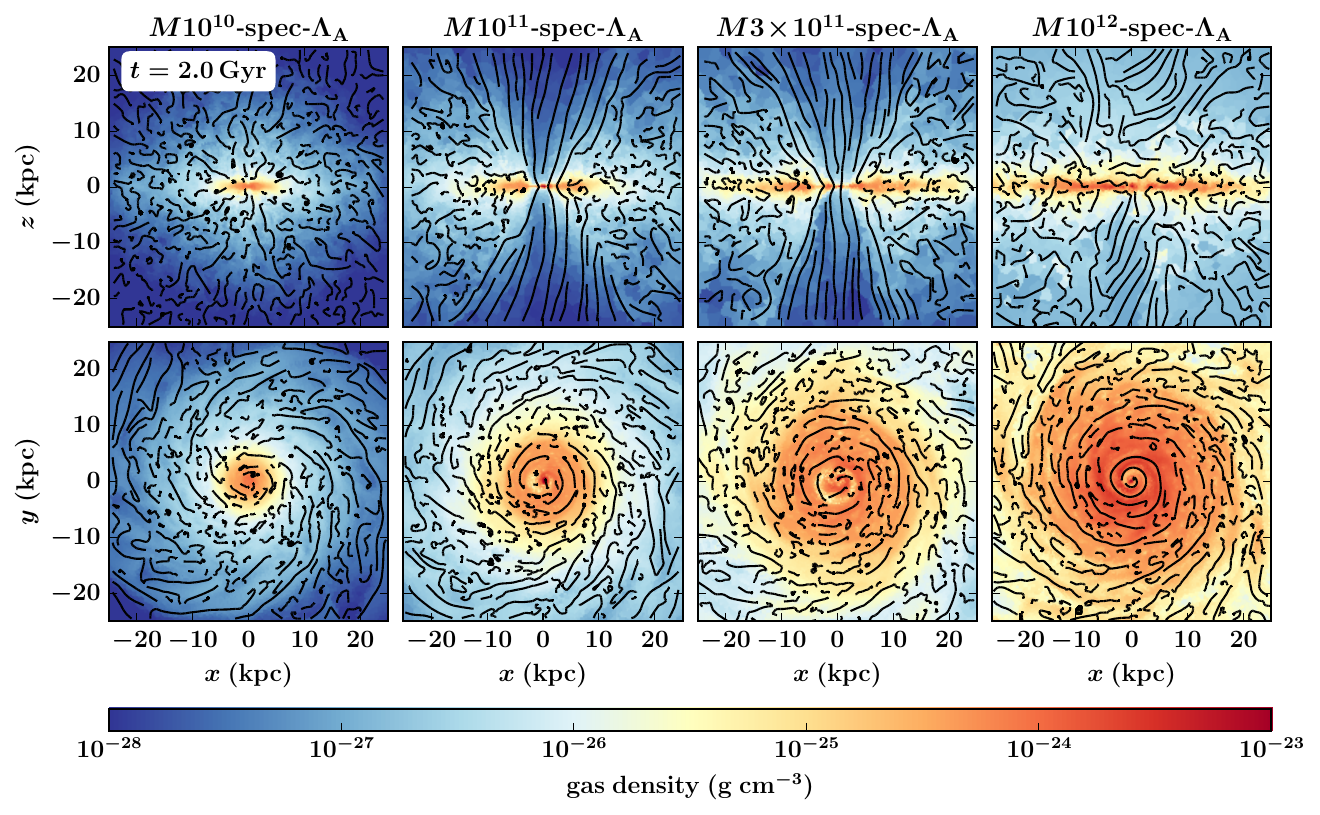}\\
    \caption{Density cuts through the centre of the galaxy (edge-on, top panels and face-on, bottom panels) and overplotted magnetic field lines for the spectral model of each halo mass at $t=2\,\Gyr$. The disc shows small scale features in the magnetic field in all cases, while the outflow cones in the two intermediate mass haloes and the CGM region in the Milky-Way mass model develop homogeneous fields with a larger coherence length.}
    \label{fig:magnetic-field-stream-lines}
    \end{minipage}
\end{figure*}

We illustrate the magnetic morphology via magnetic field lines for all spectral models in Fig.~\ref{fig:magnetic-field-stream-lines}. Depicted is the gas density color coded with overplotted streamlines. While being predominantly toroidal, the fields in the disc show more small scale features in comparison to the CGM and the outflow region, except for the smallest halo mass. The variations between different CR models for each halo mass are subdominant.

\bsp	
\label{lastpage}
\end{document}